\documentclass[a4paper,11pt]{article}
\pdfoutput=1 

\usepackage{jheppub} 

\usepackage[T1]{fontenc} 

\usepackage{amsmath,amssymb,amsfonts,amsthm,amscd,mathrsfs}
\usepackage{esvect}
\usepackage{xcolor}
\definecolor{darkblue}{rgb}{0.1,0.1,.7}
\usepackage[]{version}
\usepackage[]{graphicx}
\usepackage[]{latexsym}
\usepackage[all,cmtip]{xy}

\usepackage{ifthen}
\usepackage{soul}
\usepackage{tikz}
\usepackage{array,setspace,mathrsfs,amsfonts,yfonts,dsfont,bbm,colonequals}
\usepackage{dsfont}
\usepackage{xspace}
\usepackage{empheq}
\usepackage{extarrows}
\usepackage{subcaption}
\usepackage{enumitem}
\usepackage{xcolor}
\usepackage{hyperref}
\hypersetup{colorlinks,allcolors=blue}
\DeclareUnicodeCharacter{03B5}{\ensuremath{\epsilon}}

\numberwithin{equation}{section}

\newcommand{\be}{\begin{equation}}
\newcommand{\ee}{\end{equation}}
\newcommand{\bea}{\begin{eqnarray}}
\newcommand{\eea}{\end{eqnarray}}
\newcommand{\ba}{\begin{equation}\begin{aligned}}
\newcommand{\ea}{\end{aligned}\end{equation}}

\newcommand{\nn}{\nonumber \\}
\newcommand{\Df}{{\Delta_\phi}}

\def\g{\gamma}

\def\a{\alpha}
\def\e{\epsilon}

\def\b{\beta}
\def\d{\delta}
\def\f{\phi}
\def\t{\theta}
\def\D{\Delta}
\def\G{\Gamma}
\def\l{\lambda}

\def\ta{\tau}

\def\G{\Gamma}

\def\w{\omega}

\def\st{{\tilde{S}}}
\def\xv{{\Vec{x}}}

\newcommand{\PD}[1]{{\color{red} \bf [PD: #1]}}

\usepackage{tikz}
\tikzset{
  vtx/.style={
    circle,
    draw=blue,
    fill=blue,
    inner sep=1pt
  },
  wcirc/.style={
    circle,
    draw=white,
    fill=white,
    inner sep=2pt
  },
  bcirc/.style={
    rectangle,
    draw=black,
    fill=red,
    inner sep=1.5pt
  },
  dcirc/.style={
    circle,
    draw=black,
    fill=black,
    inner sep=1pt
  },
  rcirc/.style={
    circle,
    draw=red,
    fill=red,
    inner sep=1pt
  },
  phi/.style={
    thick
  },
  sigma/.style={
    thick,
    dashed
  },
  vl1/.style={
    thick,
    blue
  },
  vl2/.style={
    thick,
    dashed,
    blue
  },
  valign/.style={
    baseline={([yshift=-.55ex]current bounding box.center)}
  }
}

\title{\bf Bootstrapping conformal defects  on a line}
\title{\bf Bootstrapping defect operators on a line--I: Perturbation theory}
\title{\bf Bootstrapping conformal defect operators  on a line}

\author[a]{Parijat Dey } 
\author[b]{and Kausik Ghosh}

 \affiliation[a]{\small
			{  
    Department of Astrophysics and High Energy Physics,\\
S.N. Bose National Centre for Basic Sciences,\\
Salt Lake, Kolkata 700106, India}}

 \affiliation[b]{\small{Laboratoire de Physique Th\'eorique, \\ de l'\'Ecole Normale Sup\'erieure, PSL University,\\ CNRS, Sorbonne Universit\'es, UPMC Univ. Paris 06\\ 24 rue Lhomond, 75231 Paris Cedex 05, France
			}}

\emailAdd{parijat.dey@bose.res.in} 
\emailAdd{kau.rock91@gmail.com}

\abstract{
We study a conformal field theory with cubic anisotropic symmetry in  presence of a line defect. 
We compute the correlators of the low lying defect operators using Feynman diagrams and derive explicit expressions for the two, three and four point defect correlators at the cubic fixed point in $4-\e$ dimensions to $O(\e)$. We also compute the defect $g$-function for this setup and demonstrate that this is in agreement with the  $g$-theorem, which states that the $g$-function is monotonic under the renormalisation group flow  along the defect. Next, we  focus on conformal bootstrap techniques to determine the CFT data associated with the defect operators, which is the main objective of the paper. 
We utilize the framework of  crossing symmetric  Polyakov bootstrap and compute the averaged CFT data to $O(\epsilon)$ up to a finite number of ambiguities. 
We comment on unmixing the CFT data for the double trace operators  at $O(\epsilon)$ and use this to compute the $O(\epsilon^2)$ data. Finally, we study these defect correlators non-perturbatively using numerical methods and isolate them near the free theory limit close to four dimensions.
}
\begin{document}
\maketitle

\section{Introduction}
Conformal field theories (CFTs) are  sub-class of quantum field theories (QFTs) and play an important role in understanding the physics of critical phenomena. CFTs lie at the endpoints of the renormalisation (RG) group flows and characterize  the ultraviolet and infrared fixed points of QFTs. CFTs posses an enhanced symmetry which includes translation, rotation, scale invariance and special conformal symmetry. These symmetries are powerful tools to constrain the structure of the correlation functions which encode the information of the observables in the theory.  However, in order to connect with real-world systems it is important to consider the presence of impurities or defects in CFTs. Defects are used to probe the CFTs and  capture the finite size  effects in conformal systems. Defects  appear in different branches of condensed matter and statistical physics, namely the Kondo problem that arises due to magnetic impurities in metals and surface phase transitions occurring in systems with differently-ordered regions separated by domain walls are  some examples of real-world defect systems \cite{Wilson:1974sk, Affleck:1995ge,sachdev2000quantum, Vojta_2000, Chang:2018iay}. The study of CFTs in presence of defects has seen a resurgence of interest over the past few years \cite{
Cuomo:2021rkm,Cuomo:2021kfm,Sato:2021eqo, Cuomo:2022xgw, Rodriguez-Gomez:2022gbz, Giombi:2022vnz, Giombi:2023dqs, Trepanier:2023tvb, Brax:2023goj,Cogburn:2023xzw}.

Defects break some of the symmetries of the bulk CFT. Hence the correlation functions in defect theories are less constrained than those without the defects. However, defects preserve enough symmetries of the bulk CFT which provide important information about the observables. The idea of defect conformal bootstrap is to use the symmetry and consistency conditions to constrain the correlation functions of the theory. In  presence of a defect, the CFT can have a){ bulk operators}: that live on the bulk and b){ defect operators}: that are localised on the defect. This allows the correlation functions to be decomposed in different channels depending on the  operator product expansion (OPE). Crossing symmetry of the correlator imposes non-trivial constraints on the bulk and the defect CFT data, that capture the dynamical information of the CFT. These constraints can be studied using the techniques of  boundary and defect conformal bootstrap \cite{Billo:2016cpy,Liendo:2012hy,Billo:2013jda,Gaiotto:2013nva,Gliozzi:2015qsa,Gadde:2016fbj, Bissi:2018mcq,Andrei:2018die,Dey:2020lwp,Dey:2020jlc,Bianchi:2021snj,Gimenez-Grau:2022czc, Bianchi:2023gkk, Sakkas:2024dvm,Nishioka:2022odm,Nagoya:2023zky}. This approach is based on symmetry and is independent of the Lagrangian description.

Defect CFTs can as well be studied from the Lagrangian point of view using the  tools of perturbative quantum field theory and renormalisation group. 
For our purpose, we restrict our attention to one-dimensional line defects. Line defects act as a relevant perturbation in the bulk CFT and undergo an RG flow in the space of coupling. These defects are represented by a relevant line operator of scaling dimension $\D < 1$ in the UV. The entire setup is captured by an RG flow from one defect CFT in the UV to another defect CFT in the IR, during which the bulk CFT remains at the same critical point \cite{Allais:2014fqa}. The RG monotones can be constructed for such defect RG flows. It was shown in {\cite{Affleck:1991tk}} that one can define a defect $g$-function which captures universal information about the defect CFT 
and has a monotonicity property as one flows from the  UV to IR \cite{Zhou:2023fqu, Kobayashi:2018lil,Nishioka:2021uef}. The irreversibility of $g$ shows that $g$ decreases monotonically along the RG flow i.e. $g_{UV}> g_{IR}$. 
For specific types of defects this quantity can also  be connected to quantum information theory as discussed in \cite{Casini:2016fgb, Casini:2022bsu}.

The model considered in this paper is  a defect CFT with global symmetry, specifically a CFT with $N$ scalar fields $\f_i,i=1,2,\cdots N$, with a cubic anisotropic quartic interaction in the bulk in $4-\e$ space-time dimensions. These theories  are relevant to study the magnetic systems where the $O(N)$ symmetry is broken by the crystal structure of the materials. Such systems are described by  deforming the $O(N)$ interaction term by an interaction that prefers the direction of the magnetization \cite{kln}. Note that the cubic anisotropic system reduces to the Ising-like system when $N=2$. Renormalisation group analysis shows that these theories can have different bulk fixed points depending on the bulk couplings \cite{kln}. We focus on the cubic fixed point in the paper. As a next step, we insert a relevant line defect to this bulk theory which breaks the cubic symmetry. The line defect can be thought of as a localised impurity embedded in the cubic anisotropic system and plays the role of an external field. 
The line operator undergoes an RG flow  that takes it to a defect fixed point while the bulk theory remains at the cubic fixed point. We analyse the properties of the defect operators at the defect fixed point.  The cubic anisotropic model with global symmetry in the presence of line defects has been studied earlier in \cite{Pannell:2023pwz} with symmetry breaking defects  along several internal directions\footnote{Bulk theory in presence of cubic anisotropic symmetry was explored using the  conformal bootstrap approach in \cite{Dey:2016mcs, Stergiou:2018gjj,Kousvos:2018rhl, Kousvos:2019hgc}. }. 
Our bulk theory is the same as that studied in \cite{Pannell:2023pwz} whereas the defect in \cite{Pannell:2023pwz} is distinct from ours, which correspond to activating external fields along several internal directions. 

The paper aims to study the correlation functions of the defect operators  using different methods. The first part of the paper is based on computing the defect correlators using Feynman diagrams. The second part is focused on extracting the defect CFT data using the bootstrap method which relies on the conformal block decomposition of the one dimensional CFT correlator and is agnostic to the Lagrangian description of the theory. The dynamical information is encoded in the four point defect correlators. The absense of the { lightcone} limit makes the study of the one dimensional CFT correlators challenging. It can be studied efficiently using the crossing symmetric inversion formula \cite{Mazac:2018qmi} or  equivalently the Polyakov bootstrap \cite{Mazac:2018ycv,Ferrero:2019luz,Paulos:2019fkw,Paulos:2020zxx}. The idea of Polyakov bootstrap was initiated in higher dimensions in \cite{Gopakumar:2016wkt} and subsequently developed in \cite{Mazac:2016qev,Mazac:2018ycv, Gopakumar:2018xqi, Gopakumar:2021dvg,Bissi:2022mrs}\footnote{Also see \cite{Mazac:2019shk, Caron-Huot:2020adz,Penedones:2019tng, Carmi:2020ekr} for an equivalent formulation of these sum rules but using only two channel crossing symmetry. The Regge-bounded Polyakov blocks can only be made two channel crossing symmetric in higher dimensions because of the presence of spins$\geq 2$. A different category of analytic functionals, more suitable for numerical applications in higher-dimensional conformal field theories (CFTs), was studied in \cite{Paulos:2019gtx,Ghosh:2023onl}.}. This formulation is manifestly crossing symmetric and leads to efficient computation of the CFT data in  perturbative CFTs by supressing the contributions from the double trace operators in the spectrum. Similar supressions have also been observed in \cite{Bissi:2019kkx}.  We show how to compute the defect CFT data upto a finite number of unfixed parameters using this framework. At this level we combine the results from the Feynman diagram computations with  bootstrap to fix these parameters.

The main results are summarised below:
\begin{itemize}
    \item The defect fixed point is computed at $O(\e^2)$ and the $g$- function of the cubic anisotropic theory is studied perturbatively in the parameter $\epsilon$. It is shown in \eqref{guvir}   that the $g$-function decreases monotonically along the defect RG flow at $O(\e)$ satisfying $g_{UV} >g_{IR}$.
    \item 
    The correlation functions of defect operators are computed using the Feynman diagrams. There could be different types of defect operators that transform differently under the global symmetry. We denote these operators by $\f_1(\tau) $ and $\f_{\hat{a}}(\tau)$ where $\hat{a}=2,3, \cdots N $. The two, three and four point correlation functions of these operators are computed up to $O(\e)$ in \eqref{ppdefect}, \eqref{ppp3pt}, \eqref{11114pt} and \eqref{hhhh4pt}. It is observed that the operator $\f_{\hat{a}}(\tau)$ has non-zero anomalous dimension unlike the {\it tilt} operator that appears in the $O(N)$ symmetry breaking defect \cite{Cuomo:2021kfm} and transform in the vector representation of $O(N-1)$. 
    \item 
    The defect operators are  studied using the framework of Polyakov bootstrap. 
    The defect CFT data, namely the averaged scaling dimensions and the OPE coefficients are computed at $O(\e)$, upto a finite number of unfixed parameters. At this level we  combine the results from the Feynman diagram computations with  bootstrap to fix these unknown parameters. We have explicitly done this analysis for cubic  and $O(N)$ symmetric CFTs. We address  the issue of mixing of double trace operators at $O(\e)$. We  show how to disentangle the CFT data of approximate two towers at $O(\epsilon)$ by considering different correlators of two defect operators. Finally, a part of the spectrum at $O(\e^2)$ has been computed in \eqref{eqepsilon2}.
    
   
    \item 
 We do a numerical bootstrap analysis by considering the correlators of $\f_{\hat{a}}(\tau)$ when the bulk CFT is invariant under O(N) global symmetry. 
 We show that Polyakov blocks can isolate the defect CFT by considering a specific gap optimization problem in the singlet sector when $\epsilon$ is very small.
\end{itemize}

The paper is organised as follows. We begin by introducing the cubic anisotropic model  in  Sec  \ref{sec1} following \cite{kln}. The effect of the line defect in this model is discussed in Sec \ref{sec2}. In particular, we compute the beta function of the defect coupling when the bulk theory is tuned to the cubic fixed point in $4-\e$ dimensions in \eqref{defectfp}. 
In Sec \ref{gfnsec} we study the $g$-function for this set up and comment on its monotonicity property under the RG flow. We compute the correlation functions of some low-lying defect operators in Sec \ref{correlatorsec} using Feynman diagrams upto $O(\e)$. Sec \ref{bootstrapdefect} is dedicated towards the study of defect correlators using the conformal bootstrap approach. The numerical analysis is done in Sec \ref{numerics}. Finally we conclude in Sec \ref{concsec}. The appendices contain some computational details.

\section{Cubic anisotropic model}\label{sec1}
We consider a theory of $N$ scalar fields $\f_i$  $( i=1,2,\cdots N )$ with two types of $\f^4$ interaction terms in $d+1$ space-time dimensions. The action in this theory is given by 
\bea\label{action0}
S=\int d\tau\,d^{d}\xv \left[\frac{1}{2} (\partial \f_i)^2+\frac{1}{4!}\sum_{i=1}^N \l_{ijkl}\f^i \f^j \f^k \f^l\right] \,,
\eea
where we denote the coordinates as  $x^\mu \equiv (\ta, \xv)$. The interaction  vertex can be written as  \cite{kln}
\begin{align}\label{cubcoupling}
    \l_{ijkl}=\frac{g_1}{3}(\d_{ij}\d_{kl}+\d_{ik}\d_{jl}+\d_{il}\d_{jk})+g_2\,\d_{ijkl}\,,
\end{align}
where 
   \[
    \d_{ijkl}= 
\begin{cases}
    1,& \text{if } i=j=k=l\\
    0,              & \text{otherwise}\,.
\end{cases}
\]
There are two coupling constants $g_1$ and $g_2$ in \eqref{cubcoupling}. The interaction term $g_2$ breaks the $O(N)$ symmetry of the theory and is referred  to as the cubic anisotropic interaction, which is added to the $O(N)$ interaction term $g_1$.  For $g_1 \neq 0, g_2 =0$ this reduces to the theory with $O(N)$ global symmetry. Depending on the values of the coupling constants, the cubic anisotropic system can have the following four fixed points :  
\begin{itemize}
\item Gaussian fixed point: $g_1=0=g_2$\,,
\item
Ising fixed point: $g_1=0,\, g_2 \neq 0$\,,
\item
$O(N)$ fixed point: $g_2=0,\, g_1 \neq 0$\,,
\item
Cubic fixed point: $g_1\neq 0,\, g_2 \neq 0$\,.
\end{itemize}
We will focus on the cubic fixed point in what follows. The propagator in the free theory is given by \cite{Allais:2014fqa}
\begin{align}\label{2prop}
\langle\f_i(\tau, \xv) \f_j (0,\Vec{0})\rangle &=\d_{ij}\int \frac{d^d\Vec{k}\, dw}{(2\pi)^{d+1}} \frac{e^{-i(k.\xv+\w \tau)}}{\w^2+k^2}\nn
&=\d_{ij}\frac{\st_{d+1}}{(\xv^2+\ta^2)^{(d-1)/2}}\,, 
\end{align}
where
\begin{align}
\st_{d}=\frac{S_{d,2}}{(2\pi)^d}\,,\quad S_{d,a}=\frac{2^{d-a}\pi^{d/2}\G((d-a)/2)}{\G(a/2)}\,.
\end{align}
The theory is studied using dimensional regularisation in $d=3-\e$ dimensions within the minimal subtraction scheme. We quote below the values of the couplings at the cubic fixed point  in this scheme \cite{kln}
\begin{align}\label{bulkfp}
\bar{g}_{1*} &= \frac{\e}{N}+\frac{125N-19N^2-106}{27N^3}\e^2+O(\e^3)\,,\nn
\bar{g}_{2*} &=\frac{N-4}{3N}\e+ \frac{17N^3+93N^2-534N+424}{81N^3}\e^2+O(\e^3)\,.
\end{align}
where we have introduced  modified coupling constants
\begin{align}
\bar{g}_i \equiv \frac{g_i}{(4\pi)^2}\,,\quad i=1,2\,.
\end{align}
We will rename $\bar{g}_i$ as $g_i$ and drop the superscript $\Bar{g}$ in what follows.

\section{Cubic anisotropic model with a line defect}\label{sec2}
In this section we discuss the effect of adding a  line defect to \eqref{action0}.  We choose the  line defect along the field with internal direction $\f_1$ (see \cite{Allais:2014fqa, Cuomo:2021kfm} for similar analysis in the context of $O(N)$ model) and localised in the $\xv=0$ space. Since we single out a particular internal direction $`1`$ for the defect, it breaks the symmetry of the bulk theory \eqref{action0}.  The defect action is given by
\bea\label{action}
S_{\mathcal{D}}=\int d\ta\,d^{d}\xv  \left[\frac{1}{2} (\partial \f)^2+\frac{1}{4!}\sum_{i=1}^N \l_{ijkl}\f^i \f^j \f^k \f^l\right]+\g_0\int_{\mathcal{D}} d\tau \,\phi_1(\ta, \xv=0)
\eea
where $\mathcal{D}$ is the worldline of the defect. Line defects of more general type have been considered in \cite{Pannell:2023pwz} where the defects break the symmetry in different directions in the internal symmetry space. 
This line defect triggers a renormalisation group flow along the defect. Assuming that the bulk theory is at the cubic fixed point, we study the effect of the line defect as we move from the ultraviolet (UV) region to the infrared (IR).  Note that the defect operator $\f_1$ is a relevant perturbation in $d<3$  and  irrelevant in $d>3$. The defect term can be thought of as a magnetic field  breaking the cubic anisotropic symmetry with  the defect coupling 
\begin{align}
\g_{i} =   \g_0\, \d_{i 1}\,.
\end{align}
\subsection{Defect renormalisation group flow}\label{defectfp}
\begin{figure}[h!]
\begin{minipage}{0.3\textwidth}
\centering
\begin{tikzpicture}[scale=0.6]
\draw[thick] (0,0)--(4,0) ;
\node at (2,-0.5) {$(a)$};
\filldraw[black] (4,0) circle (2pt); 
\draw[thick] (0,-2)--(2,-2);
\filldraw[black] (2,-2) circle (2pt) ;
\draw[thick] (2,-2)--(4,-1);
\filldraw[black] (4,-1) circle (2pt) ;
\draw[thick] (2,-2)--(4,-2);
\filldraw[black] (4,-2) circle (2pt) ;
\draw[thick] (2,-2)--(4,-3);
\filldraw[black] (4,-3) circle (2pt) ;
\filldraw[red] (1.9,-1.9) rectangle (2.1,-2.1) ;
\node at (1.9,-2.5) {$(b)$};
\end{tikzpicture}
\end{minipage}
\begin{minipage}{0.2\textwidth}
\centering
\begin{tikzpicture}[scale=0.6]
\draw[thick] (0,-4)--(1,-4);
\filldraw[red] (0.9,-3.9) rectangle (1.1,-4.1) ;
\filldraw[red] (2.9,-3.9) rectangle (3.1,-4.1) ;
\draw[thick] (1.1,-4)--(2.9,-4);
\draw[thick] (3.1,-4)--(4.5,-4);
\filldraw[black] (4.5,-4) circle (2pt) ;
\draw[thick] (3.05,-3.9) arc (2:177:1);
\draw[thick] (3.05,-4.1) arc (-4:-175:1);
    \node at (2,-5.4) {$(c)$};
\end{tikzpicture}
\end{minipage}
\begin{minipage}{0.2\textwidth}
\begin{tikzpicture}[scale=0.5]
\draw[thick] (0,0)--(2,0) ;
\node at (3,-3) {$(d)$};
\filldraw[red] (1.9,-0.1) rectangle (2.1,0.1); 
\draw[thick] (2,0.1)--(6,2) ;
\draw[thick] (4,-2)--(6,-1) ;
\draw[thick] (4,-2)--(6,-3) ;
\filldraw[black] (6,-1) circle (3pt) ;
\filldraw[black] (6,-3) circle (3pt) ;
\filldraw[black] (6,2) circle (3pt) ;
\draw[thick] (2,0) arc (160:290:1.6);
\filldraw[red] (3.9,-1.9) rectangle (4.1,-2.1) ;
\draw[thick] (2,0) arc (292:160:-1.6);
\end{tikzpicture}
\end{minipage}
\begin{minipage}{0.2\textwidth}
\centering
\begin{tikzpicture}[scale=0.5]
\draw[thick] (0,0)--(2,0) ;
\node at (3,-3) {$(e)$};
\filldraw[red] (1.9,-0.1) rectangle (2.1,0.1); 
\draw[thick] (2,0)--(4,-2) ;
\draw[thick] (2,0)--(6,0) ;
\draw[thick] (2,0)--(6,2) ;
\draw[thick] (4,-2)--(6,-1) ;
\draw[thick] (4,-2)--(6,-2) ;
\draw[thick] (4,-2)--(6,-3) ;
\filldraw[black] (6,-1) circle (3pt) ;
\filldraw[black] (6,0) circle (3pt) ;
\filldraw[black] (6,-2) circle (3pt) ;
\filldraw[black] (6,-3) circle (3pt) ;
\filldraw[black] (6,2) circle (3pt) ;
\filldraw[red] (3.9,-1.9) rectangle (4.1,-2.1) ;
\end{tikzpicture}
\end{minipage}
\caption{Feynman digrams for $\langle \f_1(\tau, \xv=0)\rangle$ upto $O(g^2_1,g^2_2,g_1 g_2)$. The filled red square denotes the bulk couplings whereas the filled circle denotes the defect coupling. }
\label{fig:diagrams}
\end{figure}
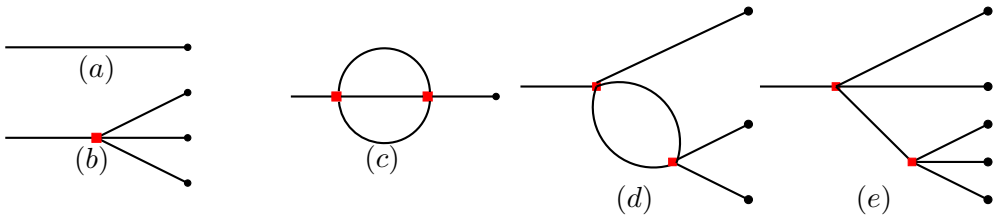

In  order to study the renormalisation group flow in the presence of the defect we need to renormalise the defect coupling $ \g_0$. We compute the beta function of the defect coupling by demanding that the renormalised one point functon $\langle \phi_1(\tau=0,{\xv})\rangle $ is finite at a distance $|\xv|$ from the defect in the limit $\e \rightarrow 0$ for all values of $g_1,g_2$ and $\g_0$. The diagrams contributing to the one point function of $\langle \phi_1(\tau=0,{\xv})\rangle $ upto  quadratic order in the bulk couplings are shown in Fig \ref{fig:diagrams}.

Note that we have not assumed that the defect coupling $\g_0$ is small. So we need to keep all the diagrams with aribitrary number of defect insertions for $O({g_1}^2,{g}^2_2, g_1\,g_2)$. We define the following renormalisation constants that relate the bare  fields to the renormalised ones in the dimensional regularisation
\begin{align}\label{recoup}
\f_i &=\frac{{\f_i}^B}{\sqrt{Z}}\, \quad, 
g^B_1 = \mu^\e\frac{ Z_{g_1}}{Z^2}\,g_1\,\quad,
g^B_2= \mu^\e \frac{Z_{g_2}}{Z^2}\,g_2, \,\quad
\g^B_0= \mu^{\e/2}\frac{ Z_{\g}}{\sqrt{Z}}\,\g\,,
\end{align}
where $\mu$ is the energy scale. We use the superscript $B$ to denote the bare quantities. The renormalisation constants for the bulk theory are unaffected by the defect and are given in terms of the renormalised couplings following \cite{kln}
\begin{align}
Z&=1-{g_1^2}\frac{N+2}{36 \epsilon }-{g_1 g_2}\frac{1}{6 \epsilon }-{ g^2_2}\frac{1}{12\epsilon }+O(g^3)\,,
\end{align}
\begin{align}
Z_{g_2}&=1+{g_1}\frac{4}{\e}+{g_2}\frac{3}{\e}+{g_1^2}\left(\frac{2N+40}{3 \epsilon^2 }-\frac{14+N}{3\e}\right)+{g_1 g_2}\left(\frac{22}{\e^2}-\frac{8}{\epsilon }\right)\nn & +\frac{g^2_2}{(4 \pi )^4}\left(\frac{9}{\e^2}-\frac{3}{\e}\right)+O(g^3)\,,\nn
Z_{g_1}&=1+{g_1}\frac{N+8}{3\e}+{g_2}\frac{2}{\e}+{g_1^2}\left(\frac{(N+8)^2}{9 \epsilon^2 }-\frac{22+5N}{9\e}\right)\nn& +{g_1 g_2}\left(\frac{12+N}{\e^2}-\frac{4}{\epsilon }\right)+{g^2_2}\left(\frac{5}{\e^2}-\frac{1}{\e}\right)+O(g^3)\,.
\end{align}
In order to evaluate the diagrams in Fig \ref{fig:diagrams} we transform the one-point function from position space to momentum space. 
This results in the following expression
\begin{align}\label{oneptk}
\langle \f_1(k) \rangle= \frac{1}{\sqrt{Z}}\left[(a)+(b)+(c)+(d)+(e)+(b')+(c')+(d')+(e')+(c'')+(d'')+(e'')\right]\,,
\end{align}
where
\begin{align}
(a)&=-\frac{\gamma^B_0}{k^2}\,,\\
(b)&=g^B_1 (\gamma^B_0)^3 \frac{2^{-2 d-1}    \Gamma (3-d) \Gamma^3 \left(\frac{d}{2}-1\right)}{3 \pi ^{d} k^{8-2 d} \Gamma \left(\frac{3 d}{2}-3\right)}\,,\\
(c) &=-\gamma^B_0 (g^B_1)^2 \,\frac{N+2}{2^{2 d+3}}\,\frac{\Gamma (2-d) \Gamma^3 \left(\frac{d-1}{2}\right)}{9 \pi ^{d+1} k^{8-2 d} \Gamma \left(\frac{3 (d-1)}{2}\right)}\,,\\
(d) &=(\gamma^B_0)^3 (g^B_1)^2\frac{(N+8)}{2^{7d-10} }\frac{\pi ^{3-\frac{3 d}{2}}  \cot (\pi  d) \sec \left(\frac{3 \pi  d}{2}\right) \Gamma (2 d-5)}{9  k^{11-3 d}\Gamma \left(\frac{d-1}{2}\right) \Gamma (d-1) \Gamma \left(2 d-\frac{9}{2}\right)}\,,\\
(e)&=(\gamma^B_0)^5 (g^B_1)^2\, \frac{2^{-4 d-1} \pi ^{-2 d} \Gamma (6-2 d) \Gamma^5 \left(\frac{d}{2}-1\right)}{3 k^{14-4 d} (d-3) (3 d-8) \Gamma \left(\frac{5 d}{2}-6\right)}\,,
\end{align}
\begin{align}
(b')&=(b)|_{g_1 \rightarrow g_2}\,,\qquad
(c')=\frac{3}{N+2} (c)|_{g_1 \rightarrow g_2}\,,\qquad
(d') = \frac{9}{N+8}(d)|_{g_1 \rightarrow g_2}\,,\qquad\nn
(e') &=(e)|_{g_1 \rightarrow g_2}\,,\qquad
(c'')=\frac{6}{N+2} (c)|_{g_1^2 \rightarrow g_1 g_2}\,,\qquad
(d'')=2(d)|_{g^2_1 \rightarrow g_1 g_2}\,,\nn
(e'')&=2(e)|_{g^2_1 \rightarrow g_1 g_2}\,.
\end{align}
We can express \eqref{oneptk} in terms of the renormalised couplings given in \eqref{recoup} and expand the expression in powers of $g_1$ and $g_2$, but not $\g$. Demanding that all poles in $\e$ should cancel at each order in $g_1,g_2$ we get the following expression for $Z_{\g}$ in the minimal subtraction scheme
\begin{align}
Z_{\g}&=1+(g_1+g_2) \frac{\gamma^2}{192 \pi ^2}\frac{1}{\e}+g_1^2\left(-\frac{\gamma ^2 \left(9 \gamma ^2+4 N+32\right)}{110592 \pi ^4\,\e}+\frac{\gamma ^2 \left(9 \gamma ^2+8 N+64\right)}{221184 \pi ^4\,\e^2}\right)\nn
&+g_2^2\left(-\frac{\gamma ^2 \left(\gamma ^2+4\right)}{12288 \pi ^4\,\e}+\frac{\gamma ^2 \left(\gamma ^2+8\right)}{24576 \pi ^4\,\e^2}\right)+g_1\,g_2 \left(-\frac{\gamma ^2 \left(\gamma ^2+4\right)}{6144 \pi ^4\,\e}+\frac{\gamma ^2 \left(\gamma ^2+8\right)}{12288 \pi ^4\,\e^2}\right)+O(g^3)\,.
\end{align}
We have determined all the renormalisation constants in the defect theory upto quadratic order in bulk coupling. We are now in a position to  compute the beta function associated with this defect. We note that since the bare coupling $\g_0$ should be independent of the energy scale we must have
\begin{align}
\mu \frac{\partial}{\partial \mu}\g^B_0= \mu \frac{\partial}{\partial \mu}\left(\mu^{\e/2}\frac{ Z_{\g}}{\sqrt{Z}}\,\g\,
 \right)=0\,.
\end{align}
This can rewritten in the form
\begin{align}
\frac{\e}{2}+\mu \frac{\partial}{\partial \mu}\log \frac{ Z_{\g}}{\sqrt{Z}}+\frac{1}{\g}\beta(\g)=0\, ,\quad {\rm where } \,\quad \beta(\g)=\mu \frac{\partial \g}{\partial \mu}\,.
\end{align}
This results in the following beta function for the defect
\begin{align}
\beta(\g)&=-\frac{\e}{2} \gamma  +(g_1+g_2)\,\frac{\gamma ^3}{96 \pi ^2}-g_1 g_2 \frac{\gamma  \left(\gamma ^4+3 \gamma ^2-1\right)}{1536 \pi ^4}\nn
&+g^2_1 \frac{\gamma  \left(-3 \gamma ^4-\gamma ^2 (N+8)+N+2\right)}{9216 \pi ^4}-g^2_2 \frac{\gamma  \left(\gamma ^4+3 \gamma ^2-1\right)}{3072 \pi ^4}+O(g^3)\,.
\end{align}
Now inserting the couplings at  bulk fixed point  \eqref{bulkfp} we get the defect fixed point 
\begin{align}\label{defectfp}
{{{\bar{\g}}}^2_*}&= \frac{ 9N}{N-1} +\frac{12032 \pi ^4 N^2+432 \pi ^2 N^2-N^2-27136 \pi ^4 N-N+54272 \pi ^4+2  }{864  (N-1) N} \epsilon+O(\e^2)\,,
\end{align}
with the redefined defect coupling 
\begin{align}
\bar{\g}\equiv \frac{{{\g}}}{4\pi^2}\,.
\end{align}
We will rename $\bar{\g}$ as $\g$ and drop the superscript in what follows. The defect fixed point has been computed earlier for cubic anisotropic symmetry with multiple line defects at leading order  in \cite{Pannell:2023pwz}. To the best of our knowledge, our result at $O(\e)$ is a new finding.
The anomalous dimension of the defect operator $\f_1$ comes out to be
\begin{align}\label{anmdimdefect}
\D({\f_1})&=1+\frac{\partial}{\partial \g}\b(\g)|_{\g_*}\nn
&=1+\e+ \e^2\bigg(-\frac{3}{2}-\frac{(N-1) (N+2)}{13824 \pi ^4 N^2}\bigg)+O(\e^3)\,.
\end{align}
This completes the study of the defect RG flow for the cubic anisotropic symmetry at $O(\e^2)$.

\subsection{Defect $g$-function}\label{gfnsec}
In this section we compute the defect $g$ function of \eqref{action}.The defect $g$-function with a circular defect of radius $R$ is defined as \cite{Cuomo:2021rkm, Cuomo:2021kfm}
\begin{align}
\log g =\log \bigg(\frac{Z^{\rm{bulk+defect}}}{Z^{\rm{bulk}}}\bigg)\,,
\end{align}
where $Z$ is the partition function of the corresponding theory. We assume that the theory is at the bulk critical point but the defect coupling in generic can be arbitrary. The $g$-function upto one-loop in the bulk couplings reads
\begin{align}\label{gfn1d}
\log g = \frac{\g^2}{2}\int_D d\ta_1  d\ta_2 G_0(x(\ta_1),x(\ta_2))-\frac{\g^4 (g_1+g_2)}{4!}\int d^d y \bigg(\int_D d\ta G_0(y,x(\ta))\bigg)^4 + O(g^2)\,.
\end{align}
We parametrize the circular defect of radius $R$ as
\begin{align}
x^{\mu}(\t)=(R \cos\t, R \sin\t,0,0,, \cdots) \,, \quad 0\leq \t \leq 2\pi\,.
\end{align}
The free propagator  \eqref{2prop} in this parametrization is given by
\begin{align}
G_0(x(\ta_1),x(\ta_2))&=\frac{\st_{d+1}}{{(x(\ta_1)-x(\ta_2)})^{d-1}}\nn
&=\frac{\st_{d+1}}{\left(4R^2 \sin^2\frac{\t_1-\t_2}{2}\right)^{(d-1)/2}}
\end{align}
and can be used to evaluate \eqref{gfn1d}. The first integral in \eqref{gfn1d} reads
\begin{align}
\int_D d\ta_1 \int_D d\ta_2 G_0(x(\ta_1),x(\ta_2))&=R^2  \st_{d+1}  \int_0^{2\pi} d\t_1 \int_0^{2\pi} d\t_2 \left(4R^2 \sin^2\frac{\t_1-\t_2}{2}\right)^{-(d-1)/2}\nn
&=-\frac{\epsilon }{4}+O(\epsilon^2)
\end{align}
This integral can be evaluated using the master integral  in \cite{Beccaria:2017rbe} resulting in the following $g$-function
\begin{align}\label{gfn1}
\log g \bigg|_{\g_*}&=-\frac{\epsilon }{8}\g^2_*+O(\epsilon^2)\nn
&=-\epsilon \frac{18 \pi ^2 N}{N-1}+O(\epsilon^2)\,.
\end{align}
The $g$-function without the defect reads
\begin{align}\label{gfn0}
\log g \bigg|_{\g=0}=1\,.
\end{align}
Comparing \eqref{gfn1} and \eqref{gfn0} we get
\begin{align}\label{guvir}
\log g \bigg|_{\g=0} >\log g \bigg|_{\g_*}\,\quad
 \Rightarrow g_{UV}  > g_{IR}\,.
\end{align}
This result is in agreement with the fact that the defect $g$-function is monotonic \cite{Cuomo:2021rkm}.


\section{Computing defect correlators using Feynman diagrams}\label{correlatorsec}
In this section we  compute the correlators of the low lying defect operators using Feynman diagrams. Using this we are able to extract the anomalous dimensions and OPE coeffcients of the low lying defect operators to $O(\e)$.
\subsection{Two point correlators}
We begin by evaluating the two point correlators of the defect theory. We first study the correlators $\langle \f_a(\tau)\f_b(0)\rangle$. This can be obtained by evaluating the following Feynman diagrams:
\begin{align}
\langle \f_a(\tau)\f_b(0)\rangle\;\; = \;\;
\begin{tikzpicture}[baseline,valign]
  \draw[thick] (-0.4, 0) -- (1.4, 0);
  \draw[dashed] (-0.2, 0) to[out=90,in=90] (1.2, 0);
\end{tikzpicture}
\;\; + \;\;
\begin{tikzpicture}[baseline,valign]
  \draw[thick] (0.0, 0) -- (2.0, 0);
  \draw[dashed] ( 0.6, 0) -- (1, 0.7) -- (1.4, 0);
  \draw[dashed] ( 0.85, 0) -- (1, 0.7) -- (1.15, 0);
  \node at (1.0, 0.70) [bcirc] {};
  \node at (0.85, 0.00) [dcirc] {};
  \node at (1.15, 0.00) [dcirc] {};
\end{tikzpicture}
\;\; + \;\; O(\e^2)\;\;.
\label{pp1}
 \end{align}
It is possible to have two types of fields on the defects depending on how they transform under the global symmetry group. We have correlators of two different fields  given by
\begin{align}
\f_1(\xv,0)\,, \qquad \f_{\hat{a}}, \qquad \hat{a}=2,3,\cdots N\,.
\end{align}
We will use $a=1, \hat{a} $ to denote these indices in what follows. The correlator can be written in terms of the renormalised fields
\begin{equation}
    \phi_1=\sqrt{{Z}_{\phi_1}} [\phi_1],\,\,\,  \phi_{\hat{a}}={\sqrt{Z}_{t_{\hat{a}}}}[t_{\hat{a}}] .
\end{equation}
We refer to the operators $t_{\hat{a}}$ as anomalous `tilt' operators. Their dimensions remain protected if the bulk symmetry possesses a conserved current, behaving similarly to typical tilt operators. However, in the cubic anisotropic case there is no conserved current and therefore we will see below that it has anomalous dimensions.

The renormalisation factors are obtained by demanding the finiteness of the two point functions $\langle \f_a (\tau)\f_b(0)\rangle$ which results in
\begin{align}
{Z}_{\phi_1}&=1-\frac{\g^2}{32 \pi^2 \e}(g_1+g_2)+O(g^2_1, g^2_2, g_1 g_2)\,,\nn
{Z}_{t_{\hat{a}}}&= 1-\frac{\g^2}{96\pi^2 \e}g_1+O(g^2_1, g^2_2, g_1 g_2)\,.
\end{align}
The defect two point function takes the following form
\begin{align}\label{ppdefect}
\langle [\phi_1](\tau) [\phi_1](0)\rangle &=\frac{\mathcal{N}_\f}{{\tau}^{2 \D_{\f_1}}}\,,\nn
\langle[t_{\hat{a}}](\tau) [t_{\hat{b}}](0)\rangle &=\d_{\hat{a} \hat{b}} \frac{\mathcal{N}_t}{{\tau}^{2 \D_t}}\,,
\end{align}
where
\begin{align}\label{2ptanmdim}
\D_{\f_1} &=1+\e+O(\e^2)\,,\nn
\mathcal{N}_\f &= \frac{1}{4 \pi ^2}\left[ 1-\frac{\e}{2}(3+2\log \pi+2\g_E)\right]+O(\e^2)\,,\nn
\D_t &=1-\e \frac{(N-4)  }{2 (N-1)}+O(\e^2)\,,\nn
\mathcal{N}_t &= \frac{1}{4 \pi ^2}\left[1+\frac{\e (N-4)}{2(N-1)}\,\bigg(\frac{3}{4-N}+\g_E+\log \pi\bigg)\right]+O(\e^2)\,,
\end{align}
and $\g_E$ is the Euler Gamma function. Note that both the operators $\f_1$ and $t_{\hat{a}}$ receive corrections at $O(\e)$. The fact that the operator $t_{\hat{a}}$ has anomalous dimension is unlike the defect in the $O(N)$ symmetric model, where the operators $t_{\hat{a}}$ are protected. In the $O(N)$ model, the operator $t_{\hat{a}}$ quantifies the breaking of the Noether current on the defect whereas for the cubic symmetry group, there is no such conserved current in the bulk.


\subsection{Three point correlators }
In this section we determine the three point defect correlators or the OPE coefficients. These results are obtained from Feynman diagram computations, keeping terms up to $O(\epsilon)$ in the $\epsilon$ expansion. Symmetry fixes the functional form of the three-point correlators 
\begin{align}
\langle \mathcal{O}_1 (\tau_1)  \mathcal{O}_2 (\tau_2)  \mathcal{O}_3(\tau_3)\rangle &=\frac{\mathcal{N}_{ \mathcal{O}_1}\mathcal{N}_{ \mathcal{O}_2}\mathcal{N}_{ \mathcal{O}_3} C_{\mathcal{O}_1 \mathcal{O}_2\mathcal{O}_3}}{\tau_{12}^{\D_1+\D_2-\D_3}\tau_{13}^{\D_1+\D_3-\D_2}\tau_{23}^{\D_2+\D_3-\D_1}}\,.
\end{align}
We focus on the simplest three point functions with only the fundamental scalar. It is obtained by the following Feynman diagram at $O(\e)$
\begin{align}
 \langle \, \phi_a (\tau_1) \, \phi_b(\tau_2) \, \phi_c(\tau_3) \, \rangle
 \;\; = \;\;
 \begin{tikzpicture}[baseline,valign]
  \draw[thick] (0, 0) -- (1.4, 0);
  \draw[dashed] (0.2, 0) to[out=90,in=90] (1.2, 0);
  \draw[dashed] (0.5, 0) -- (0.7, 0.25);
  \draw[dashed] (0.9, 0) -- (0.7, 0.25);
  \node at (0.7, 0.28) [bcirc] {};
  \node at (0.9, 0.00) [dcirc] {};
\end{tikzpicture}
 \;\; + \;\; O(\e^2)\,.
 \label{eq:diag-3pt-PPP}
\end{align}
Since this diagram involves one bulk vertex, this is proportional to $g_{1*}, g_{2*} \propto \e$ it is sufficient to evaluate this integral in $d=4$ \cite{Gimenez-Grau:2022czc}.
\begin{align}
\lim_{\xv_1,\xv_2,\xv_3,\xv_4\rightarrow 0} \int  \frac{d\tau_4 d^4x_5}{x^2_{15}x^2_{25}x^2_{35}x^2_{45}}=\frac{2\pi^4}{\tau_{12}\tau_{13}\tau_{23}}\,.
\end{align}
The OPE coefficient takes the following form
\begin{align}
C_{\f_a\f_b\f_c} \propto \frac{g_{1*} \gamma_{*}}{3} \left(\d_{ab} \d_{c1}+\d_{a1}\d_{bc}+\d_{ac}\d_{b1}\right)+g_{2*}\gamma_{*}\,\delta_{abc1}\,.
\end{align}
This can be decomposed into irreducible representations of the global cubic symmetry for $(N-1)$ scalar fields. This results in the following non-vanishing OPE coefficients
\begin{align}\label{ppp3pt}
C_{\f_1\f_1\f_1} &= \sqrt{\frac{N-1}{N}}\pi\e+O(\e^2)\,,\nn 
C_{\f_{\hat{a}}\f_{\hat{b}}\f_1} &= \frac{\pi \e}{\sqrt{N(N-1)}}+O(\e^2)\,.
\end{align}

\subsection{Four point correlators }
In this section we study the  four point correlator in the defect theory \eqref{action}.  
The four point correlator upto $O(\e)$ is obtained by summing over the following Feynman diagrams
\begin{align}
\langle \f_a(\tau_1) \f_b(\tau_2) \f_c(\tau_3) \f_d(\tau_4)\rangle &=\;
\begin{tikzpicture}[baseline,valign]
  \draw[thick] (-0.4, 0) -- (1.4, 0);
  \draw[dashed] (-0.2, 0) to[out=90,in=90] (0.4, 0);
  \draw[dashed] (0.6, 0) to[out=90,in=90] (1.2, 0);
\end{tikzpicture}
\;+\;
 \begin{tikzpicture}[baseline,valign]
  \draw[thick] (-0.4, 0) -- (1.4, 0);
  \draw[dashed] (-0.2, 0) -- (0.5, 0.7) -- (1.2, 0);
  \draw[dashed] ( 0.3, 0) -- (0.5, 0.7) -- (0.7, 0);
  \node at (0.5, 0.7) [bcirc] {};
\end{tikzpicture}
\;+\;
\begin{tikzpicture}[baseline,valign]
  \draw[thick] (-0.4, 0) -- (1.6, 0);
  \draw[dashed] (-0.2, 0) to[out=90,in=90] (0.4, 0);
  \draw[dashed] ( 0.6, 0) -- (1, 0.7) -- (1.4, 0);
  \draw[dashed] ( 0.85, 0) -- (1, 0.7) -- (1.15, 0);
  \node at (1.0, 0.70) [bcirc] {};
  \node at (0.85, 0.00) [dcirc] {};
  \node at (1.15, 0.00) [dcirc] {};
\end{tikzpicture}
\;+\; O(\e^2)\,.
\end{align}
We can write the renormalised correlator in terms of the cross-ratio
\begin{align}
z=\frac{\tau_{12}\tau_{34}}{\tau_{13}\tau_{24}}\,.
\end{align}
The correlator of the operator $\f_1$ reads
\begin{align}\label{11114pt}
&\langle [\f_1](\tau_1) [\f_1](\tau_2) [\f_1](\tau_3) [\f_1](\tau_4)\rangle \nn &=\frac{1}{(\tau_{12}\tau_{34}) ^{2\D_{\f_1} }}\bigg[1+z ^{2\D_{\f_1}}+\left(\frac{z }{z -1}\right)^{2\D_{\f_1}}+\e \frac{4 (N-1)  }{N}I(z)\bigg]+O(\e^2)\,,
\end{align}
where 
\begin{align}
I(z)&=\frac{z ^2 \log z }{1-z }+z  \log (1-z )\,.
\end{align}

The correlator of the $t_{\hat{a}}$ operator can be written as
\begin{align}\label{hhhh4pt}
&\langle[t_{\hat{a}}](\tau_1) [t_{\hat{b}}](\tau_2) [t_{\hat{c}}](\tau_3) [t_{\hat{d}}](\tau_4)\rangle \nn &= 
\frac{1}{{(\tau_{12}\tau_{34})^{2\Delta_t}}}\bigg[{\mathcal{G}_S } \delta _{\hat{a} \hat{b}} \delta _{\hat{c} \hat{d}}+  \mathcal{G}_T \left(\delta _{\hat{a} \hat{c}} \delta _{\hat{b} \hat{d}}+\delta _{\hat{a} \hat{d}} \delta _{\hat{b} \hat{c}}-2 \delta _{\hat{a} \hat{b} \hat{c} \hat{d}}\right)+  \mathcal{G}_V\left(\delta _{\hat{a} \hat{b} \hat{c} \hat{d}}-\frac{\delta _{\hat{a} \hat{b}} \delta _{\hat{c} \hat{d}}}{{N-1}}\right)\nn &+ \mathcal{G}_A\left(\delta _{\hat{a} \hat{d}} \delta _{\hat{b} \hat{c}}-\delta _{\hat{a} \hat{c}} \delta _{\hat{b} \hat{d}}\right)\bigg]+O(\e^2)\,,
\end{align}
where $\Delta_t$ is defined in \eqref{2ptanmdim}.
We have decomposed the correlator into the irreducible representations of the cubic  anisotropic symmetry  for $N-1$ scalar fields \cite{kln,Dey:2016mcs}. This ensures that the defect breaks the global symmetry in one internal direction but keeps it invariant in the remaining $N-1$ directions. The functions in \eqref{hhhh4pt} are defined below
\begin{align} \label{4ptcubic}
{\mathcal{G}_S }&=\bigg(1+\frac{1}{N-1}\frac{z^2}{(z-1)^2}\left(2-2 z+z^2\right)+\epsilon \lambda^S_1 \log z ^2 + \epsilon  \lambda^S_2 \log (z -1)^2\bigg)\,,\nn
 \lambda^S_1 &=-\frac{z^2}{2(N-1)^2N(z-1)^2}\bigg(\left(z^2+6 z -6\right) N^2-4 \left(z^2+3 z -3\right) N+12 (z -1)\bigg)\,,\nn
 \lambda^S_2 &=\frac{z}{2(N-1)^2 N(z-1)^2}\bigg(8 z ^2 (N-1)^2+((28-15 N) N-16) z +8 (N-1)^2-4(N-1)(1-z)^2\bigg)\,,\nn
 {\mathcal{G}_T }&=\bigg( \frac{z ^2 \left(z ^2-2 \xi +2\right)}{2 (z -1)^2}+\e\, \lambda^T_1 \log z ^2 + \epsilon  \lambda^T_2 \log (z -1)^2\bigg)\,,\nn
  \lambda^T_1 &= -\frac{z ^2}{4 (N-1) N (z -1)^2}\bigg(((z -2) z +2) N^2-4 (z -2)^2 N-8 z +8\bigg)\,,\nn
  \lambda^T_2 &= \frac{z }{4 (N-1) N (z -1)^2}\bigg(8 (N-1) z ^2+(N-20) N z +16 z +8 N-8\bigg)\,,\nn
 {\mathcal{G}_V }&=\bigg( \frac{z ^2 \left(z ^2-2 z +2\right)}{(z -1)^2}+\e\, \lambda^V_1 \log z ^2 + \epsilon  \lambda^V_2 \log ( z-1)^2\bigg)\,,\nn
 \lambda^V_1 &= \frac{z ^2}{2 (N-1) N (z -1)^2}\bigg((2-z  (z +2)) N^2+4 \left(z ^2+z -1\right) N-8 z +8\bigg)\,,\nn
  \lambda^V_2 &= \frac{z }{2 (N-1) N (z -1)^2}\bigg(4 z ^2 N^2-7 z  N^2+4 N^2+8 z ^2-12 z ^2 N+20 z  N-12 N-16 z +8\bigg)\,,\nn
{\mathcal{G}_A }&= \bigg(-\frac{(z -2) z ^3}{2 (z -1)^2}+\e\,\lambda^A_1\log z ^2 + \epsilon  \lambda^A_2 \log (z -1)^2 \bigg)\,,\nn
 \lambda^A_2 &=\frac{(N-4) z ^2}{4 (N-1) (z -1)^2}\,,\nn
 \lambda^A_1 &=\lambda^A_2 (z -2) z  \,.
\end{align}
Note that the correlators computed in \eqref{11114pt} and \eqref{hhhh4pt} are crossing symmetric under the exchange of $x_2 \leftrightarrow x_4$ which serves as a consistency check.

\section{Bootstrapping four-point correlators}\label{bootstrapdefect}
In this section, we use analytic bootstrap tools  to compute the CFT data, namely the scaling dimensions and the OPE coefficients  of the defect correlators. The goal is to show that the CFT data can be fixed upto a finite number of unfixed parameters using the analytic functionals\footnote{While analytic functionals may encompass a wide range of functionals, for our purposes, we specifically consider those that equivalently lead to Polyakov bootstrap equations. Therefore, we use the term Polyakov bootstrap or analytic functionals interchangeably without distinction.  } \cite{Mazac:2018ycv,Ferrero:2019luz, Ghosh:2021ruh}. 
 These parameters as well as the correlators can be fixed unambiguously if we use the inputs from other techniques like the Feynman diagrams. Let us first consider the four-point correlator of $\phi_1$ on the defect. On the line defect the correlators are invariant under $SL(2,\mathbb{R})$ : the one-dimensional conformal group. Therefore the four-point correlator is fixed upto an arbitrary function of single cross ratio $z$,
\begin{equation}
    \langle \phi_1(\tau_1)\phi_1(\tau_2)\phi_1(\tau_3)\phi_1(\tau_4) \rangle =\frac{1}{\tau_{12}^{2\Delta_{\phi_1}}\tau_{34}^{2\Delta_{\phi_1}}}\mathcal{G}(z).
\end{equation}
These functions $\mathcal{G}(z)$ can be expanded in terms of $SL(2,\mathbb{R})$ conformal blocks,
\begin{equation} \label{cbexp}
    \mathcal{G}(z)=\sum_{\Delta}a_{\Delta} G_{\Delta}(z),
\end{equation}
where 
\begin{align}
G_{\Delta}(z)=z^{\Delta}\, _2F_1(\Delta,\Delta,2\Delta;z)\,,
\end{align}
are the conformal blocks and $a_{\Delta}$ are the OPE coefficients of the exchanged operators with scaling dimensions $\D$. It was shown in \cite{Mazac:2018ycv} that $\mathcal{G}(z)$ also admits an expansion in terms of crossing symmetric Polyakov blocks, $\mathcal{P}_{\Delta}(z)$ with the same OPE coefficients $a_{\Delta}$,

\be \label{polyexp}
 \mathcal{G}(z)=\sum_{\Delta}a_{\Delta} \mathcal{P}_{\Delta}(z)\,.
\ee
Polyakov blocks admit an expansion in terms of the $SL(2,\mathbb{R})$ blocks and their derivatives,
\be
\mathcal{P}_{\Delta}(z)=G_{\Delta}(z)-\sum_n \alpha_n(\Delta) G_{\Delta_n}(z) -\sum_n \beta_n(\Delta) \partial G_{\Delta_n}(z).
\ee
If we sum over the Polyakov blocks after multiplying with appropriate OPE coefficients and claim that the two expansions, eqn \eqref{cbexp} and \eqref{polyexp} agree, we get the following two sets of discrete sum rules that constrain the CFT data,
\be \label{booteqn}
\begin{split}
&\sum_{\Delta} a_{\Delta} \beta_n(\Delta)=0,\, \forall\,{ \text{non-negative}} \,\, n,  \\
& \sum_{\Delta} a_{\Delta} \alpha_n(\Delta)=0,\, \forall\, \text{non-negative} \,\, n. 
\end{split}
\ee
Here $\a_n(\Delta)$ and $\beta_n(\Delta)$ are the linear functionals that can be computed as contour integrals of crossing symmetric equation with kernels constrained appropriately \cite{Mazac:2018ycv}. We have briefly reviewed this construction and the relation with exchange and contact Witten diagrams in appendix \ref{appendix:rev}. For our purpose what is important is that $\alpha_n(\Delta)$ and $\beta_n(\Delta)$ are kinematically fixed functions of $\Delta$ and the OPE data of any correlator of identical scalar operators on a line have to satisfy these sum rules. We now apply these sum rules to compute the CFT data. The functionals can be written as follows,
\be
\begin{split}
& \beta_n(\Delta)=\mathcal{N}(\Delta,\Delta_{\phi})\left(b^s_n(\Delta,\Delta_{\phi})+2 b^t_n(\Delta,\Delta_{\phi})\right)+\lambda R_n,\\
& \alpha_n(\Delta)=\mathcal{N}(\Delta,\Delta_{\phi})\left(a^s_n(\Delta,\Delta_{\phi})+2 a^t_n(\Delta,\Delta_{\phi})\right)+\lambda S_n\,.
\end{split}
\ee
 The parameter $\lambda$ is fixed by demanding that one of the functional equation is missing,i.e., say we decide to use $n=0$ beta equation to fix it then,
\begin{equation}\label{fixinglambda}
\begin{split}
    & \mathcal{N}(\Delta,\Delta_{\phi})\left(b^s_0(\Delta,\Delta_{\phi})+2 b^t_0(\Delta,\Delta_{\phi})\right)+\lambda R_0=0,\\
  &  \implies \lambda=-\frac{\mathcal{N}(\Delta,\Delta_{\phi})}{R_0}\left(b^s_0(\Delta,\Delta_{\phi})+2 b^t_0(\Delta,\Delta_{\phi})\right).
    \end{split}
\end{equation}
All the explicit expressions of relevant quantity in above equations are given in Appendix \ref{app:expression}.
Now we have to assume the spectrum and the bootstrap equations \eqref{booteqn} will fix the CFT data. We assume that upto first sub-leading order in perturbation theory only the identity and the double trace operators appear in the OPE of $\phi_1$ \footnote{A similar type of computation appeared in \cite{Mazac:2018ycv} where the authors studied $\phi^4$ flow in $AdS_2$. Here the spectrum is more subtle as there is a mixing of double trace operators and that would change the details of the computation. } at the leading order in any perturbative theory,
\be\label{opelead}
\phi_1 \times \phi_1\sim I+(\phi_1\phi_1)_n+\cdots\,.
\ee
Note  that for CFTs with $N$-scalar fields transforming under  a global symmetry with a line defect, there is also another family of double trace operators $(\phi_a\phi^a)_n$, $ a=2,3,\cdots N$, having the same bare dimension as $\f_1$, which mix with the one written above in \ref{opelead}. Hence, our result contains averaged values with contributions of multiple operators that are degenerate at lower orders in the perturbation theory. Later on, we will show how to disentangle the individual CFT data in section \ref{unmix}.
The CFT data of double trace operators can be expanded in the parameter $\epsilon$,
\be
\begin{split}
& \Delta_n=2+2n+\gamma^1_n \epsilon+ O(\e^2)\,,\\
& a_n=a^0_n+a^1_n \epsilon+O(\e^2)\,,
\end{split}
\ee
for $n=0,1,2,\cdots$.
 The dimension of the external scalar $\Delta_{\phi_1}$ is also taken as an input,
 \begin{equation}
     \Delta_{\phi_1}=1+\epsilon+O( \epsilon^2)+\cdots\,.
 \end{equation}
 The leading order OPE coefficients can be obtained by looking at $\alpha_n(\Delta)$ equation \eqref{booteqn}. The OPE coefficient is given by,
 \begin{equation}
     a^0_n=-\alpha_n(0)=\frac{2 \Gamma^2 (2 n+2) \Gamma (2 n+3)}{\Gamma (2 n+1) \Gamma (4 n+3)}.
 \end{equation}
 Then we expand $\beta_n(\Delta)$ and  find the following two contributions at $O(\epsilon)$:
 \begin{enumerate}[label=\roman*.]
 \item
 contribution of the operator with dimension $\Delta_0$,
 \begin{equation}\label{e11}
   \epsilon  \frac{a^0_0 (\gamma^1_0-2)  \Gamma^3 (n+1) \Gamma \left(n+\frac{3}{2}\right)}{2^{2 n} (n!)^2 \Gamma \left(2 n+\frac{3}{2}\right)},
 \end{equation}

\item
 contribution of the operator with dimension $\Delta_n$,
\begin{equation}\label{e12}
-\epsilon \,a^0_n (-2 + \gamma^1_n)\,.
\end{equation}
\end{enumerate}
Summing over the above two contributions \ref{e11} and \ref{e12} we find the following expressions for the averaged value of the anomalous dimensions of the double trace operators 
\begin{align}
\langle a^0_n \gamma^1_n \rangle=\frac{2 \Gamma (2 n+2)^2 (\langle a^0_0 \gamma^1_0\rangle+4 n (2 n+3))}{\Gamma (4 n+3)}\,.
\end{align}
Similarly, we expand $\alpha_n(\Delta)$ to find the following corrections to the OPE coefficients 
\be
\begin{split}
& \langle a^1_n \rangle=-\frac{\sqrt{\pi } 2^{-4 n-1} (\langle a^0_0 \gamma^1_0\rangle-4) \Gamma (2 n+2) \left(-H_{2 n+1}+H_{2 n+\frac{1}{2}}+\log (4)\right)}{\Gamma \left(2 n+\frac{3}{2}\right)}\\
&+\frac{8 \left(H_{2 n+1}+H_{2 n+2}-H_{4 n+2}-1\right) \Gamma (2 n+2)^2 \Gamma (2 n+3)}{\Gamma (2 n+1) \Gamma (4 n+3)}\,,
\end{split}
\ee
where $H_n$ is the Harmonic number of order $n$
\begin{align}
H_n = \sum_{k=1}^{n}\frac{1}{k}\,.
\end{align}
Thus the CFT data to order $O(\epsilon)$ is fixed in terms of one unknown parameter, $a_0 \gamma^1_0$, which is the CFT data of the leading double trace operator with $n=0$. This serves as an independent cross-check of our results derived from Feynman diagrams \eqref{11114pt}. For convenience we provide the value of the unfixed parameter below,
\be
\begin{split}
& \langle a_0\gamma^1_0\rangle=\frac{2 (2 N+19)}{N+8},\,\,\,\,\,\,\,\text{for O(N),}\\
& \langle a_0\gamma^1_0\rangle=\frac{4 (2 N-1)}{N},\,\,\,\,\,\,\,\text{for cubic.}
\end{split}
\ee
\subsection{Four point correlator of anomalous tilt operator}
In our defect setup, the fields $\phi_{\hat{a}}$ with $\hat{a}=2,3,...,N$ transform under the cubic anisotropy group. The correlators can  be expanded into the irreducible sectors of the global symmetry group,
\begin{equation}
    G^{ijkl}(z)=\sum_c T^c_{ij,kl} G_c(z)\,,
\end{equation}
where the sum runs over all channels which in this case are singlet (S), symmetric vector (T), traceless (V) and antisymmetric (A) representation of cubic symmetry group. Each channel can be expanded in terms of conformal blocks or equivalently Polyakov blocks,
\begin{equation}
G_c(z)=\sum_{\Delta} a^c_\Delta G_{\Delta}(z)=\sum_b \sum_{\Delta} a^b_\Delta P^{c|b}_{\Delta}(z).
\end{equation}
$b$ runs over all the irreducible sectors. A general prescription for any real group was given in \cite{Ghosh:2021ruh}. For the present case, there are three Regge bounded contact Witten diagrams. Therefore the Polyakov blocks will be defined in terms of exchange Witten diagrams and by suitably adding these contact Witten diagrams (see Appendix \ref{app:contact}). This will lead to losing three equations and therefore our perturbative solutions constructed using this basis will be fixed in terms of three parameters (ambiguities). Since these functionals didn't appear anywhere We have written down the explicit form of the sum rules in appendix \ref{appendix:rev}.

While matching this data with prediction from Feynman diagram we have to provide three CFT data to fix the ambiguities which bootstrap alone can't fix. Now as usual first we assume that upto $O(\epsilon)$ there are only identity and double trace like operator in the OPE of $t_{\hat{a}}$ and $t_{\hat{b}}$. Therefore there will be identity in the singlet ($S$) channel and then in the $S$,$T$,$V$ channel the operators will have dimension $2+2n+ \gamma^O \epsilon$ and in the antisymmetric($A$) channel the operator will have dimension $3+2n+\gamma^A_n \e$. We provide the OPE data of these sectors,
\small{
\begin{equation}
    \begin{split}
      & \langle a^S_n\gamma^S_n \rangle=  \frac{\sqrt{\pi } 2^{-4 n-1} (10 n (2 n+3) \langle a^S_1\gamma^S_1\rangle-(n-1) (2 n+5) \langle a^S_0\gamma^S_0\rangle) \Gamma (2 n+2)}{5 \Gamma \left(2 n+\frac{3}{2}\right)},\\
      &  \langle a^T_n\gamma^T_n\rangle  =-\frac{\sqrt{\pi } (N-4) 2^{-4 n-1} n (2 n+3) \Gamma (2 n+2)}{(N-2) \Gamma \left(2 n+\frac{3}{2}\right)}\\
      & +\frac{\sqrt{\pi } 4^{-2 n-1} \Gamma (2 n+2) ((N-1) n (2 n+3) (\langle a^S_0\gamma^S_0\rangle-10 \langle a^S_1\gamma^S_1\rangle)+10 (N-2) a^T_0\gamma^T_0)}{5 (N-2) \Gamma \left(2 n+\frac{3}{2}\right)},\\
      & \langle a^V_n\gamma^V_n \rangle = -\frac{\sqrt{\pi } (N-4) 2^{-4 n-1} \left(N+4 n^2+6 n+1\right) \Gamma (2 n+2)}{(N-2) \Gamma \left(2 n+\frac{3}{2}\right)}\\
      & + \frac{\sqrt{\pi } (N-1) 2^{-4 n-1} n (2 n+1) \Gamma (2 n) \left(\langle a^S_0\gamma^S_0\rangle \left(13 N+4 n^2+6 n-23\right)-10 \langle a^S_1\gamma^S_1\rangle \left(3 N+4 n^2+6 n-3\right)\right)}{5 (N-2) \Gamma \left(2 n+\frac{3}{2}\right)}\\
      & -\frac{\sqrt{\pi } (N-1) 2^{-4 n} n (2 n+1) a^T_0 \gamma^T_0 \Gamma (2 n)}{\Gamma \left(2 n+\frac{3}{2}\right)},\\
      & \langle a^A_n \gamma^A_n\rangle=\frac{\pi  (N-4) 4^{-4 n-6} \Gamma (4 n+5) \left(-4 (N-2) \Gamma (4 n+7)-\pi ^3 \Gamma (2 n+3) \Gamma (2 n+4)\right)}{(N-2) (N-1) \Gamma \left(2 n+\frac{5}{2}\right) \Gamma \left(2 n+\frac{7}{2}\right) \Gamma (4 n+4)}\\
      &+ \frac{\pi ^4 (N-1) 2^{-8 n-11} (n+1)\langle a^S_0\gamma^S_0\rangle \Gamma (2 n+3) \Gamma (2 n+4)}{5 (N-2) \Gamma \left(2 n+\frac{5}{2}\right) \Gamma \left(2 n+\frac{7}{2}\right)}-\frac{\pi ^4 (N-1) 4^{-4 n-5} (n+1) \langle a^S_1\gamma^S_1\rangle \Gamma (2 n+3) \Gamma (2 n+4)}{(N-2) \Gamma \left(2 n+\frac{5}{2}\right) \Gamma \left(2 n+\frac{7}{2}\right)}.
    \end{split}
\end{equation}}
We have obtained similar expression for OPE coefficients as well. This agrees with CFT data obtained from  \eqref{4ptcubic}. We write down the value of undetermined parameters found from Feynman diagram computation below,
\begin{equation}
    \langle a^S_0 \gamma^S_0 \rangle=\frac{6 \left(N^2-2 N+2\right)}{(N-1)^2 N},\,\,\,\,\,\,\,\,\,\langle a^S_1 \gamma^S_1 \rangle=-\frac{2 \left(N^2-7 N-3\right)}{5 (N-1)^2 N},\,\,\,\,\,\,\,\,\,\,\, a^T_0 \gamma^T_0 =-\frac{N^2-8 N+4}{(N-1) N}.
\end{equation}
\subsection{$O(N)$ global group}

Now we deal with the defect CFT when the bulk CFT has global symmetry $O(N)$. So the defect contains a tilt operator with protected dimension 1 and it transforms under the $O(N-1)$ group. We obtain the following OPE data,
\begin{equation}
\begin{split}
& \langle a^S_n \gamma^S_n \rangle\\
&=\frac{\sqrt{\pi } 2^{-4 n-1} \Gamma (2 n+2) ((N-1) \langle a^S_0 \gamma^S_0\rangle (3 (N-1)+2 (N-2) n (2 n+3))-2 (N-2) (N+1) n (2 n+3) a^T_0 \gamma^T_0)}{3 (N-1)^2 \Gamma \left(2 n+\frac{3}{2}\right)},
\end{split}
\end{equation}
{\small
\begin{equation*}
\begin{split}
& \langle a^T_n \gamma^T_n\rangle \\
&=\frac{2^{-4 n-3} }{9 \Gamma \left(2 n+\frac{3}{2}\right)} \Bigg(\frac{3\ 4^{n+2} a^T_0 \gamma^T_0 \left(3 (N-1)+2 (N+1) n^2+3 (N+1) n\right) \Gamma (n+1)^3 \Gamma \left(n+\frac{5}{2}\right)}{(N-1) (2 n+3) (n!)^2}\\
&-\frac{5 \sqrt{\pi } \langle a^S_0 \gamma^S_0\rangle\left(\frac{3}{2}\right)_{n-1} \left((2)_{n-1}\right){}^2 \left(\frac{7}{2}\right)_{n-1} \Gamma (2 (n+1)) \Gamma (2 n+3)}{(1)_{n-1} \left(\left(\frac{5}{2}\right)_{n-1}\right){}^2 (3)_{n-1} \Gamma (2 n+1)}\Bigg),\\
& \langle a^A_n \gamma^A_n\rangle =\frac{\sqrt{\pi } 2^{-4 n-3} (n+1) \Gamma (2 n+4) ((N-1) \langle a^S_0\gamma^S_0\rangle- (N+1) a^T_0\gamma^T_0)}{3 (N-1) \Gamma \left(2 n+\frac{5}{2}\right)}\,,
\end{split}
\end{equation*}}
where $(a)_b=\frac{\G(a+b)}{\G(a)}$ are the Pochhammer symbols.
We also obtain similar expressions for OPE coefficients. The CFT data obtained here is in  agreement with \cite{Gimenez-Grau:2022czc}\footnote{The tensor structure of traceless symmetric sector is related as $T^{here}_{\hat{a}\hat{b}\hat{c}\hat{d}}=2T^{there}_{\hat{a}\hat{b}\hat{c}\hat{d}}$. The convention for other sectors are same.} For convenience we also write the values of undetermined parameter found from Feynman diagram computation below,
\begin{equation}
    \langle a^S_0 \gamma^S_0 \rangle=\frac{2 (N+1)}{(N-1) (N+8)},\,\,\,\,\,\,\,\,\,\,\,\,\,\ a^T_0\gamma^T_0=\frac{1}{2}\frac{4}{N+8}.
\end{equation}

\subsection{Mixed correlators}
Recently Polyakov bootstrap was generalized to multiple correlators setup \cite{Ghosh:2023wjn}. We will use this setup for our case with a straightforward generalization to CFTs in presence of global symmetry. So here we consider mixed correlator,
\begin{equation}
    \mathcal{G}^{ijkl}(z)=\sum_{\Delta} a_{\Delta} G^{ijkl}_{\Delta}(z).
\end{equation}
It also admits an equivalent expansion in terms of crossing symmetric mixed Polyakov blocks,
\begin{equation}
    \mathcal{G}^{ijkl}(z)=\sum_{\Delta} a_{\Delta} P^{ijkl}_{\Delta}(z)
\end{equation}
The explicit sum rules can be written down in the following form,
\begin{equation}
    \sum_{\Delta,P=\pm}a_{\Delta,P} \left(r^{ij,kl;s}_{\Delta,P} \alpha^{ij,kl;s}_{I_s}+ r^{ij,kl;t}_{\Delta,P} \alpha^{ij,kl;t}_{I_s}+r^{ij,kl;u}_{\Delta,P} \alpha^{ij,kl;u}_{I_s}\right)=0.
\end{equation}
Now let us focus on the case of the correlator of the fields $\phi_1$ and $t_{\hat{a}}$ and $a$ is the vector index of a global symmetry and understand what does the above expression mean in this context. First of all, this comes from expanding the correlator in terms of a crossing symmetric object. Therefore unlike the usual $s-$channel expansion of $\langle\phi_1 \phi_1 t_{\hat{a}}t_{\hat{b}}\rangle$ where only the $s-$ channel OPE appears, here exchanges in all three channel appears in the sum rule. Moreover, we have introduced a vector $r^{ij,kl;c}$ which indicates how a particular operator appears in the $c$ channel of correlator $\langle\phi_i\phi_j\phi_k\phi_l\rangle$. In the correlator $\langle\phi_1\phi_1 t_{\hat{a}} t_{\hat{b}}\rangle$, singlets appear only in $s$-channel, therefore $r^{ij,kl;s}=1$ and $r^{ij,kl;t/u}=0$. Similarly for vector exchanges, only $t$ and $u$ channels have non-zero $r$. Strictly speaking, $r^{ij,kl;s}$ also includes the tensor structure which in this case is $\delta_{ab}$ which we have taken into account and since it is the same in all channels, therefore, we throw it out. Next, we see that functionals are labeled by $I_{s}$ which accounts for all the $s-$ channel double traces we can write down. So in our present case, we have $(\phi_1 \phi_1)_n$ and $(t_a t_b)_n$ having dimension of $2\Delta_{\phi_1}+2n$ and $2\Delta_t+2n$ and we find sum rules labelled by each of these $n$. The explicit construction of the functionals $\alpha^{ij,kl;c}$ from the relevant exchange and contact Witten diagrams was explained in \cite{Ghosh:2023lwe} so we don't repeat it here.
We find from bootstrapping $\langle\phi_1 \phi_1 t_{\hat{a}}t_{\hat{b}}\rangle$ of bulk $O(N)$  and cubic theory the average OPE times anomalous dimensions of the singlets\footnote{In actual derivation one has to deal with another subtlety that is the collision of poles. Since the dimension of $\phi_1$ and $t_a$ operator starts off as 1 so in perturbation theory the $\alpha_{(\phi_1\phi_1)_n}$ and $\alpha_{(tt)_n}$  functionals will collide to give $\alpha_n$ and $\beta_n$. The results are derived from these functionals.},
\begin{equation} \label{onmixed}
 \langle\sqrt{a_{\phi_1\phi_1O_n} a_{ttO_n}}\gamma_n\rangle=\frac{\sqrt{\pi } \sqrt{a_{\phi_1\phi_1O_0} a_{ttO_0}}\gamma_0 2^{-4 n-1} \Gamma (2 n+2)}{\Gamma \left(2 n+\frac{3}{2}\right)}\,,
\end{equation}
We also find similar expressions for correction to the product of the averaged OPE coefficient. 
The unfixed parameters are 
\be
\begin{split}
& \sqrt{a_{\phi_1\phi_1O_0} a_{ttO_0}}\gamma_0=\frac{2}{N+8},\,\,\,\,\, \text{for O(N)}\\
& \sqrt{a_{\phi_1\phi_1O_0} a_{ttO_0}}\gamma_0=\frac{4}{N}, \,\,\,\,\,\text{for cubic}.
\end{split}
\ee
\subsection{Unmixing the CFT data} \label{unmix}
In this section, we examine the OPE data of double-trace singlet operators. It's important to highlight that we can construct these operators in two different ways using the derivatives along the defect: $\phi_1 (\partial_{\tau})^{2n} \phi_1$ and $\phi_a(\partial_{\tau})^{2n} \phi^a$. Notably, there is exactly one operator at level $n$, characterized by  $2n$ derivatives of $\tau$, which corresponds to the direction along the defect. The most general operator at this level can be expressed as a linear combination of $n$ operators. However, to ensure they are primaries- meaning they are annihilated by the special conformal transformation generator $(K)$- we impose $n-1$ constraints. This leads to only one independent primary operator remaining at level n involving fields $\phi_1$ and $\phi_a$ separately. Note that if we only had these operators then considering the correlators of two external fields would be sufficient to disentangle the two towers completely.

However in our case, there is another category of double-trace operators, specifically $\phi_a \partial_i \partial^i ....\phi_a$, where $i$ represents directions orthogonal to the defects. By incorporating these orthogonal derivatives, we can construct multiple but finite number of primary operators at each level $n$. In this paper we considered correlators involving two external operators, i.e., $\phi_1$ and $t_{\hat{a}}$. So the best we can do is to isolate two towers using these correlators. At a first glance it might seem that this data is not useful as the true theory of line defects have a different spectrum. It is important to mention that the  one dimensional numerical bootstrap works very differently  than its higher dimensional counterpart. In the extremal spectrum of mixed correlator bootstrap (on the line) we can only see two towers of operators \cite{Ghosh:2023wjn}. So in the space of CFTs when we will search for this line defect numerically, we can compare the CFT data with our analytical prediction. In other words  we can only approximately be close to a true theory with finite set of correlators. In section \ref{numerics}, we will carry out this analysis using a single correlator, leaving the mixed correlator numerics for future work.



Since we have solutions for correlators involving $\phi_1$ and $ t_{\hat{a}}$ we can actually unmix the anomalous dimension of two double trace  operators each with a bare dimension of $2+2n$.  Using this data we would be able to predict $O(\epsilon^2)$ result. Basically we have six data in a mixed fashion ($\langle a^0_{\phi_1 \phi_1 O_n}\gamma^1_n \rangle$, $\langle a^0_{ttO_n}\gamma^1_n \rangle$, $\langle a^0_{\phi_1 \phi_1 O_n} \rangle$, $\langle a^0_{ttO_n}\rangle$, $\langle \sqrt{a^0_{ttO_n}a^0_{\phi_1 \phi_1O_n}}\rangle$,$\langle\sqrt{a^0_{t tO_n}a^0_{\phi_1 \phi_1O_n}} \gamma^1_n \rangle$ ). Here $O_n$ denotes a singlet operator. Since we have six parameters and six equations we can just solve for them and we find unmixed physical anomalous dimension and OPE coefficient.

For $O(N)$ theory we obtain the following data for individual double trace operators,
\begin{equation}
\begin{split}
& a^0_{\phi_1 \phi_1O_n}\\=& \frac{\Gamma (2 n+2)^2}{A \Gamma (2 n+1) \Gamma (4 n+3)}\\
&\bigg\{ \bigg(4 \left(2 n^2+3 n+1\right) \left(2 N+4 (N+8) n^2+6 (N+8) n+19\right) \Gamma (2 n+1)\\
& +\left(-3 N-4 (N+8) n^2+A-6 (N+8) n-20\right) \Gamma (2 n+3)\bigg),\\
&- \bigg(4 \left(2 n^2+3 n+1\right) \left(2 N+4 (N+8) n^2+6 (N+8) n+19\right) \Gamma (2 n+1)-(3 N+4 (N+8) n^2\\
&+A +6 (N+8) n+20) \Gamma (2 n+3)\bigg) \bigg\},\\
\end{split}
\end{equation}

\be
\begin{split}
a^0_{t tO_n}=& \frac{16^{-n}  \Gamma(2 + 2 n) }{(N-1) A \Gamma (2 n+1) \Gamma \left(2 n+\frac{3}{2}\right) \Gamma (4 n+3)}\\
&\bigg\{ \bigg(\sqrt{\pi } (N+1) \left(2 n^2+3 n+1\right) \Gamma (2 n+1) \Gamma (4 n+3)+16^n (-3 N-4 (N+8) n^2\\
& +A-6 (N+8) n-20) \Gamma (2 n+2) \Gamma (2 n+3) \Gamma \left(2 n+\frac{3}{2}\right)\bigg),\\
&\bigg(16^n \left(3 N+4 (N+8) n^2+A+6 (N+8) n+20\right) \Gamma \left(2 n+\frac{3}{2}\right) \Gamma (2 n+2) \Gamma (2 n+3) \\
& -\sqrt{\pi } (N+1) \left(2 n^2+3 n+1\right) \Gamma (2 n+1) \Gamma (4 n+3)\bigg) \bigg\},\\
\end{split}
\ee

\begin{equation}
\begin{split}
 \gamma^1_n=&\bigg\{
 \frac{\sqrt{B}+3 N+2 (N+8) n (2 n+3)+20}{2 (N+8) (n+1) (2 n+1)},\frac{-\sqrt{B}+3 N+2 (N+8) n (2 n+3)+20}{2 (N+8) (n+1) (2 n+1)}\bigg\}   
\end{split}
\end{equation}
where 
\begin{equation*}
\begin{split}
     B=& N^2 \left(16 n^4+48 n^3+44 n^2+12 n+1\right)+N \left(256 n^4+768 n^3+784 n^2+312 n+40\right)\\
     &+64 (n+1) (2 n+1) (4 n (2 n+3)+5)
     \end{split}
\end{equation*}
and
\begin{equation*}
\begin{split}
     A=& \bigg(N^2+40 N+16 (N+8)^2 n^4+48 (N+8)^2 n^3+4 \left(11 N^2+196 N+864\right) n^2,\\
     & +12 \left(N^2+26 N+144\right) n+320\bigg)^{\frac{1}{2}}.
     \end{split}
\end{equation*}

For cubic symmetry, we find the following set of unmixed data,
\begin{equation}
    \begin{split}
 \gamma^1_n=& \bigg\{ \frac{6 N^2-14 N+2 N^2 (n+1)^2-N^2 (n+1)+4 N (n+1)^2-2 N (n+1)+8}{2 (N-1) N (n+1) (2 (n+1)-1)}\\
 & +\left((-1)^{n+1}+3\right) \frac{\sqrt{C}}{4 (N-1) N (n+1) (2 (n+1)-1)},\\
 & \frac{6 N^2-14 N+2 N^2 (n+1)^2-N^2 (n+1)+4 N (n+1)^2-2 N (n+1)+8}{2 (N-1) N (n+1) (2 (n+1)-1)}\\
 & -\left((-1)^{n+1}+3\right) \frac{\sqrt{C}}{4 (N-1) N (n+1) (2 (n+1)-1)}\bigg\}
 \end{split}
\end{equation}
with the following OPE coefficients,
\begin{equation}
    \begin{split}
       a^0_{\phi_1 \phi_1O_n}=& \bigg \{ -\frac{\sqrt{\pi } 2^{-4 n} \Gamma (2 n+2) (N ((\gamma^1_n)_2+((\gamma^1_n)_2-2) n (2 n+3)-4)+2)}{N (\text{$\gamma $1}-(\gamma^1_n)_2) \Gamma \left(2 n+\frac{3}{2}\right)}\\
        & \frac{\sqrt{\pi } 2^{-4 n} \Gamma (2 n+2) (N ((\gamma^1_n)_1+((\gamma^1_n)_1-2) n (2 n+3)-4)+2)}{N ((\gamma^1_n)_1-(\gamma^1_n)_2) \Gamma \left(2 n+\frac{3}{2}\right)}\bigg\}
    \end{split}
\end{equation}
and
\begin{equation}
    \begin{split}
        a^0_{t tO_n}=& \bigg\{
-\frac{\frac{2 \gamma^1_n)_2 \Gamma (2 n+2)^2 \Gamma (2 n+3)}{(N-1) \Gamma (2 n+1) \Gamma (4 n+3)}-\frac{\sqrt{\pi } 16^{-n} (3 (N-2) N+(N-4) N (-n) (2 n+3)+6) \Gamma (2 n+2)}{(N-1)^2 N \Gamma \left(2 n+\frac{3}{2}\right)}}{\gamma^1_n)_1-\gamma^1_n)_2},\\
&\frac{\Gamma (2 n+2) \left(\frac{(\gamma^1_n)_1 (N-1) \Gamma (2 n+3)}{\Gamma (4 n+2)}+\frac{\sqrt{\pi } 16^{-n} (N (-3 N+(N-4) n (2 n+3)+6)-6)}{N \Gamma \left(2 n+\frac{3}{2}\right)}\right)}{(N-1)^2 ((\gamma^1_n)_1-(\gamma^1_n)_2)}\bigg\}
    \end{split}
\end{equation}
where we have denoted the two unmixed anomalous dimensions at each level $n$ as $(\gamma^1_n)_1$ and $(\gamma^1_n)_2$ and 
\begin{equation}
\begin{split}
 C=&\bigg(-6 N \left(3 (-1)^{n+1}-5\right) \left(2 (n+1)^2-n-1\right)\\
 & -\frac{1}{2} N^2 \left(3 (-1)^{n+1}-5\right) \left(36 (n+1)^4-36 (n+1)^3-39 (n+1)^2+24 (n+1)+1\right)\\
& \frac{1}{2} N^3 \left(3 (-1)^{n+1}-5\right) \left(36 (n+1)^4-36 (n+1)^3-21 (n+1)^2+15 (n+1)+2\right)\\
& -\frac{1}{8} N^4 \left(3 (-1)^{n+1}-5\right) \left(6 (n+1)^2-3 (n+1)-2\right)^2\bigg).
\end{split}
\end{equation}

\subsection{ $O(\epsilon^2)$ data}
Since we have disentangled the OPE data in the previous order, we can now compute the data that appears in four-point function of the $\phi_1$ operator at $O(\epsilon^2)$. To the best of our knowledge, this is a new computation. At $O(\epsilon^2)$, in addition to the double trace operators, there will be a new operator in the $\phi_1 \times \phi_1$ OPE, which is $\phi_1$ itself. We work out a specific example below. We start with the nonperturbative sum rule,
\be
\sum_{\Delta}a_{\Delta}\beta_{1}(\Delta)=0.
\ee
Then we expand this equation to $O(\e^2)$. We isolate the contribution of double twist operator with $n=0$ and $n=1$ from rest. The contribution of $n=0$ operator at this order is the following:
\be
\epsilon ^2 \left(\frac{42 R(2) a^0_0 (\gamma^1_0-2)^2}{\pi }+\frac{1}{300} a^0_0 \bigg(\gamma^1_0 (2299 \gamma^1_0-9110)+30 \gamma^2_0-60 \delta\phi_2+9024\bigg)+\frac{1}{10} a^1_0 (\gamma^1_0-2)\right),
\ee
where $R(\Delta)=\frac{1}{4} \pi  \left(\psi ^{(1)}\left(\frac{\Delta +1}{2}\right)-\psi ^{(1)}\left(\frac{\Delta }{2}\right)\right)$.
The $n=1$ double trace operator contributes,
\be
\epsilon ^2 \left(\frac{13580 R(4) a^0_1 (\gamma^1_1-2)^2}{\pi }+\frac{1}{105}a^0_1(55054 (\gamma^1_1-4)\gamma^1_1-105\gamma^2_1+210 \delta\phi_2+220216)-a^1_1 (\gamma^1_1-2)\right)
\ee
Rest of the double trace operators contribute the following,
\be
\begin{split}
&\epsilon ^2\sum_{m=2}^{\infty}  \frac{a^0_m (\gamma^1_m-2)^2 16^m  \Gamma \left(2 m+\frac{5}{2}\right) }{\pi ^{3/2} (m-1) m (2 m+3) (2 m+5) \Gamma (2 m+2)} \\
&\times \bigg(\pi  \left(2 (m-1) m (2 m+3) (2 m+5) \left(4 m^2+6 m+5\right)-1\right)+2 \pi  (m-1) m (2 m+3) (2 m+5)  \\
&(2 m (2 m+3)(m (2 m+3)+4)+7) \left(\psi ^{(1)}\left(\frac{1}{2} (2 m+3)\right)-\psi ^{(1)}\left(\frac{1}{2} (2 m+2)\right)\right)\bigg)
\end{split}
\ee
Finally the operator $\phi_1$ itself contributes,
\be
a_{\phi_1}\frac{\left(35-4 \pi ^2\right) \epsilon ^2}{12 \pi ^2}
\ee
So far what we have done is true for the real theory. We can see the appearance of quantities like $a^0_m (\gamma^1_m-2)^2$ which requires unmixing of leading order anomalous dimension. But here we use the unmixed value we have obtained in the previous section to get an estimate of approximate answer. Substituting the value of the OPE coefficient $a_{\phi_1}$ from Feynman diagram computation, we obtain \footnote{To evaluate the sum over m we evaluate the sum to a very high precision and then we write it in a basis of transcendental functions and fix the rational coefficients.} the following results for $O(N)$ defects,
\begin{equation} \label{eqepsilon2}
\begin{split}
    &\bigg\langle a^0_n \gamma^2_n+  a^1_n \gamma^1_n \bigg\rangle\bigg|_{n=1}\\ =&\frac{1}{1800 (N+8)^2}\bigg({2664 N^2+95707 N-5400 \pi ^2 (N+8)-3600 (N+8) \zeta (3)+605960}\bigg)\\
  &  +\frac{1}{10}  \bigg\langle a^0_n \gamma^2_n+  a^1_n \gamma^1_n\bigg\rangle \bigg|_{n=0}
    \end{split}
\end{equation}
We find similar expressions for other $n$ but we couldn't manage to obtain a closed form in $n$. Note this equation is fixed up to $n=0$ data which bootstrap can't fix as expected.
This is an important set of data because this captures the double discontinuity of the correlator. We also obtain similar expressions for cubic anisotropic defect,
\begin{equation}
\begin{split}
  &\bigg\langle a^0_n \gamma^2_n+  a^1_n \gamma^1_n \bigg\rangle\bigg|_{n=1}\\= &  -\frac{N^2+N-2}{691 \pi ^4 N^2}-\frac{\pi ^2 (N-1)}{3 N}+\frac{1}{N}\left({8 \zeta (3)-\frac{13571}{900}}\right)-8 \zeta (3)+\frac{14903}{900}\\
  & + \frac{1}{10}\bigg\langle a^0_n \gamma^2_n+  a^1_n \gamma^1_n\bigg\rangle \bigg|_{n=0}.
  \end{split}
\end{equation}
It will be interesting to consider larger set of external correlators so that we can unmix the larger number of double trace opertors and get an estimate of $O(\e^2)$ result of the actual underlying line defect theory.

\section{Numerical Results}\label{numerics}
In this section, we will design an optimization problem that seems to give us the defect CFTs as extremal CFT. For this purpose let's consider defect with unbroken $O(2)$ symmetry. An extensive numerical analysis using derivative functionals has been done in \cite{Gimenez-Grau:2022czc} for $\epsilon=1$. Here we focus on the small $\epsilon$ region. From the perturbation theory result, we can see that the defect theory in one dimension starts as a generalized free theory when $\epsilon$ is very small and as it is tuned to $\epsilon=1$ we go to a strongly coupled sector. Now we consider the four-point correlation function of the tilt operator with protected dimension 1. We set $\epsilon=10^{-3}$ such that we can trust the epsilon expansion results at this point. We notice that in the Polyakov basis, two functionals are missing. In other words, the deformation of GFF theory is ambiguous up to two unfixed parameters. Also, there is an operator $\phi_1$ which appears in the OPE of the tilt operator with itself at $O(\epsilon^2)$. This operator is not part of a GFF family and it must be considered as an extra degree of freedom which we need to fix from somewhere else. To summarise, there are three parameters for which we don't have equations. Hence we give inputs to fix these ambiguities and then run the numerics. First, we demand that there are operators $\Delta_S=1.001$ and $\Delta_T=2.00018$. We also set the gap in the antisymmetric sector to 3 (which is the free GFF dimension of the first nontrivial antisymmetric operator). Now as a function of $a_{ttS}$ and $a_{ttT}$  we look for a spectrum that has operator $\Delta_S$=2.001. We find that this problem is very constrained and there is a regime close to ($a_{ttT}=1, a_{tt\phi_1}=0.0001$) where we can find such an operator. All other dimensions also appear to be of the same magnitude as it is predicted for $O(N)$ defect theory from perturbative computations.

\begin{figure}[ht]
    \centering
    \begin{subfigure}{0.435\textwidth}
        \centering
        \includegraphics[width=\textwidth]{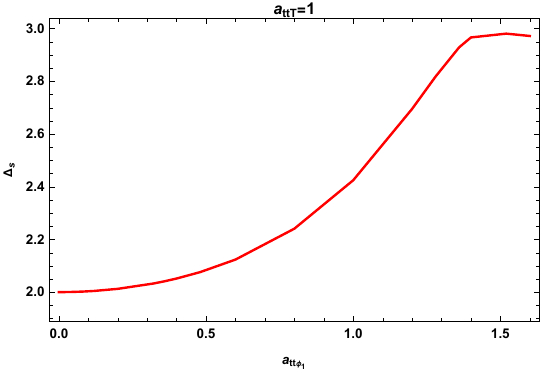}
        \caption{$\epsilon=10^{-3}$}
         \end{subfigure}
    \hfill
    \begin{subfigure}{0.435\textwidth}
        \centering
    \includegraphics[width=\textwidth]{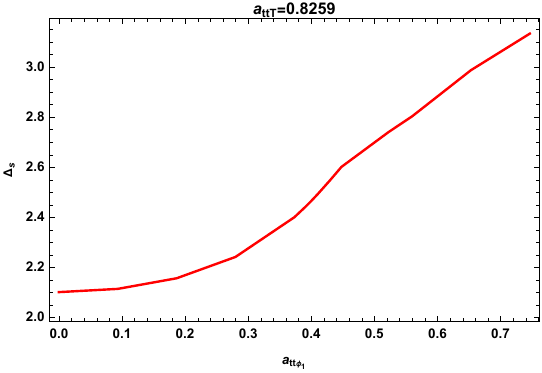}
        \caption{$\epsilon=1$.}
         \end{subfigure}
         \caption{Maximizing the gap in the singlet sector with varying the OPE coefficient $a_{tt\phi_1}$ with fixed $a_{ttT}$.}
    \label{fig: o(N)}
\end{figure}

Now we perform the same exercise but at $\epsilon=1$ with $\Delta_S=1.55$ and $\Delta_T=2.23$. Again we look for a spectrum that contains an operator with  $\Delta_S$=2.36 as a function of $a_{ttS}$ and $a_{ttT}$. Again we find an interesting point that contains such a spectrum. Note  that this point is strongly correlated.  Therefore we can't compare the dimensions of the other operators.

We observe  that by considering this single correlator we couldn't resolve the degeneracy of double trace operators. We see only one double trace-like operator in the spectrum. This is expected. Because in perturbation theory we observed that we have to consider a mixed correlator of tilt and $\phi_1$ operator to lift off the degeneracy of the double trace operators. The same phenomena is also observed in cubic anisotropic case. But in this case there are total four ambiguities even when we try to bootstrap single correlator of anomalous tilt operator. We are currently investigating mixed correlator setups for both cases using analytic functionals, which will be presented as a separate work in the near future.

\section{Discussions}\label{concsec}
We study the effect of a line defect in a CFT with cubic anisotropic symmetry in $4-\e$ dimensions. The line defect acts as a relevant deformation that breaks the bulk global symmetry and only a subgroup of the cubic symmetry is preserved depending on the direction of the defect. We work in the setup where the bulk theory is at the cubic fixed point and there is an RG flow along the defect. The correlation functions of various defect operators at the defect fixed point are computed using the Feynman diagrams at $O(\e)$. We have  computed the defect $g$-function related to the impurities introduced in the system due to the  defect and shown that this defect entropy is monotonic under the defect RG flow. 

We also bootstrap the correlators of the defect operators using the methods of Polyakov bootstrap. We find that this method determines the defect CFT data  up to a finite number of ambiguities. This outcome is expected since adjusting these parameters could lead to different  CFTs residing on a line. By using the results from the Feynman diagrams to fix these  ambiguities, we demonstrate that all other data precisely aligns with the results from the Feynman diagram computations at $O(\epsilon)$. By fine-tuning this unfixed parameter, we are able to simultaneously reproduce the data for $O(N)$ and cubic defects.  At $O(\epsilon)$, there is a mixing of double-trace operators, which we subsequently unmix for two operators. As a result, we compute the OPE data of the operators appearing in the $\phi_1 \times \phi_1$ OPE at $O(\epsilon^2)$, up to one unfixed parameter. We provide explicit expressions for both $O(N)$ and cubic anisotropic defects for $n=1$ double-trace operators and  obtain similar data for other values of $n$. Since this flow is perturbatively fixed up to a finite ambiguity we showed that it is possible to locate $O(N)$ defect CFTs by numerical bootstrap where we scan over these unfixed parameters. 

Along the line of this work, there are several interesting directions to pursue in  future. One interesting avenue would be to try fixing the {ambiguity associated to the defect CFT data} at $O(\epsilon^2)$. 
On a more ambitious ground, it would be very nice to push these results to higher orders in $\epsilon$. At $O(\epsilon^3)$ there will be new towers of operators and it would be fascinating to understand how  we can bootstrap these new data.

In this work we have studied a single correlator using the numerical bootstrap methods. It would be interesting to bootstrap mixed correlators and see how the new operators appear in the spectrum. One can also apply the fuzzy sphere technique \cite{Hu:2023ghk} to reproduce some of the low-lying spectrum of this defect theory. That can guide us to set up the optimization problem while improving the assumptions as well as help in the numerical search at $\epsilon=1$, i.e. non-perturbative regime. We hope to report on this mixed correlator setup for both $O(N)$ and cubic defect CFT in the near future. 

Another useful idea would be to incorporate the bulk locality constraints in numerics. Recently analytic functionals have been constructed that act on the form factors involving the bulk operators and boundary fields \cite{Levine:2023ywq, Meineri:2023mps}. This has led to powerful sum rules. It would be  interesting to develop such functionals tailored for our line defect setup. In contrast to the focus of the present paper, which mainly centers on the defect operators while taking the bulk data as input, these sum rules encompass the data pertaining to both bulk and defect operators. Consequently, one can reasonably anticipate a more restrictive parameter space for the CFTs. It's worth noting that we also need to formulate the sum rules involving the two-point correlation functions of the bulk stress tensor along the lines of \cite{Meineri:2023mps}. This endeavor would enable us to establish a positively semi-definite framework involving bulk and boundary OPE coefficients, facilitating their direct application in numerical bootstrap methods. We hope to return to these questions in the near future.

Furthermore, it would be intriguing to explore the broader scope of defect setups. Recently, surface defects have been investigated in a series of papers \cite{Giombi:2023dqs, Sun:2023vwy, Cuomo:2023qvp, Krishnan:2023cff}. Analyzing how these modifications affect the correlation functions of the defect operators and the parameter space of general defect CFTs would provide valuable insights in this direction.

\begin{acknowledgments}
		We thank Aleix Gimenez-Grau, Slava Rychkov, Andreas Stergiou and Philine van Vliet for useful discussions. KG thanks Miguel Paulos for various discussions and collaboration on related topics. KG is grateful to IHES for their hospitality during the final stage of this work. We thank Agnese Bissi, Aninda Sinha and Andreas Stergiou for comments on the draft. This research is co-funded by the European Union (ERC, FUNBOOTS, project number 101043588). Views and opinions expressed are however those of the author(s) only and do not necessarily reflect those of the European Union or the European Research Council. Neither the European Union nor the granting authority can be held responsible for them. 
		
	\end{acknowledgments}

\appendix
\section{Quick review of Polyakov Bootstrap} \label{appendix:rev}
In this appendix, we briefly outline the construction of analytic functionals. Let's consider the four-point correlator of an operator $\phi$ with dimension $\Delta_{\phi}$. The correlator can be expanded in terms of the conventional $SL(2,{\mathbb{R}})$  conformal blocks,
\begin{equation}
    G(z)=\sum_{\Delta} a_{\Delta} G_{\Delta}(z),
\end{equation}
or equivalently we can also expand the same correlator in terms of the Polyakov blocks,
\begin{equation}
    G(z)=\sum_{\Delta} a_{\Delta} P_{\Delta}(z).
\end{equation}
The conceptual difference between these two approaches is that the conventional blocks are not manifestly crossing symmetric, whereas the Polyakov blocks are crossing symmetric on their own. Therefore conventionally the bootstrap equations arise by demanding the crossing symmetry (equivalence) in different channel expansions. But in the second approach, the blocks are manifestly crossing symmetric and hence the bootstrap equations arise from demanding consistency with OPE expansion. This can be seen when we decompose the Polyakov blocks in $SL(2,R)$ blocks itself,
\begin{equation}
    P_{\Delta}(z)=G_{\Delta}(z)-\sum_n (\alpha_n(\Delta)(z) G_{\Delta_n}+\beta_n(\Delta) \partial G_{\Delta_n})(z).
\end{equation}
Now if we sum over OPE coefficient,
\begin{equation} \label{polyaexp}
  \sum_{\Delta} a_{\Delta}  P_{\Delta}(z)= \sum_{\Delta} a_{\Delta}G_{\Delta}(z)-\sum_n \left[ \bigg(\sum_{\Delta} a_{\Delta}\alpha_n(\Delta)\bigg)G_{\Delta_n}(z)+ \bigg(\sum_{\Delta} a_{\Delta}\beta_n(\Delta)\bigg) \partial G_{\Delta_n}(z)\right],
\end{equation}
we see that the first term will give us the full correlator but there are other terms present and this could be a valid expansion of the physical correlator only if,
\begin{equation}
\begin{split}
&    \sum_{\Delta} a_{\Delta}\alpha_n(\Delta)=0\\
 &   \sum_{\Delta} a_{\Delta} \beta_n(\Delta)=0.
    \end{split}
\end{equation}
Also, notice in equation \eqref{polyaexp} that we have actually swapped the sum over $n$ and $\Delta$ and this is a subtle step as we are commuting two infinite sums. Poyakov blocks are suitably constructed so that we are allowed to do this swapping\cite{Mazac:2018ycv,Qiao:2017lkv}. We call $\alpha_n(\Delta)$ and $\beta_n(\Delta)$ functionals because we can find them by acting on crossing anti-symmetric vector,
\begin{equation}
    F_{\Delta}(z)=z^{-2\Delta_{\phi}} G_{\Delta}(z)-(1-z)^{-2\Delta}G_{\Delta}(1-z)
\end{equation}
by integrating against its discontinuity around 1,
\begin{equation}
    \omega_n(\Delta)=\int_{1}^{\infty} dz h^{\omega}_{n}(z) \mathcal{I}m(F_{\Delta}(z)).
\end{equation}
Another property of the analytic functionals are that they satisfy certain orthogonality relations depending on the particular construction we are interested in. For solutions where $\Delta_n$, i.e. the dimensions of the conformal block that appeared in the expansion of Polyakov block is equal to $2\Delta_{\phi}+2n$, the orthogonality relations are the following,
\begin{equation}
    \begin{split}
       &  \alpha_n(\Delta_m)=\delta_{m,n}\,\,\,\,\,\,\,\,\,  \alpha_n(\partial \Delta_m)=\delta_m,n-d_n \delta_{m,0},\\
          &  \beta_n(\Delta_m)=0\,\,\,\,\,\,\,\,\,\,\,\,\,\,\,\,\,  \beta_n(\partial \Delta_m)=\delta_{m,n}-c_n \delta_{m,0}.
    \end{split}
\end{equation}
These properties fix the form of the kernel $h^{\omega}_n(z)$. The most efficient and quicker way to get these functionals is from the $AdS_2$ Witten diagrams. The sum of $s$, $t$ and $u$ exchange Witten diagrams in $AdS_2$ is crossing symmetric on its own and also Regge bounded. This crossing symmetric sum also admits an equivalent expansion in terms of $SL(2,R)$ conformal blocks,
\begin{equation}
    W^s_{\Delta}(z)+W^t_{\Delta}(z)+W^u_{\Delta}(z)=\sum_{n} \bigg[\alpha^W_n(\Delta)(z)G_{\Delta_n}(z)+\beta^W_n(\Delta)(z)\partial G_{\Delta_n}(z)\bigg].
\end{equation}

Now we can apply analytic functionals to the crossing symmetric sum of Witten diagrams and we get the following equations,
\begin{equation}
    \alpha_n(\Delta)=\alpha^W_n(\Delta)-d_n \beta^W_{0}(\Delta),
    \beta_n(\Delta)=\beta^W_n(\Delta)-c_n \beta^W_{0}(\Delta).
\end{equation}
So this tells us that the analytic functionals are related to crossing symmetric Witten diagrams up to some shifts. These shifts have a physical interpretation in terms of Regge bounded contact terms in $AdS_2$. Since these are Regge bounded contact terms therefore these deformations should be allowed by our basis. So if we had concluded that only the exchange Witten diagrams form the basis dual to the generalized free field of bosons then we would exclude the $\phi^4$ deformations in $AdS_2$ which is a perfectly well-defined theory. So the main message is that crossing symmetric sum of exchange Witten diagram and the Regge bounded contact Witten diagrams with scalar external legs forms a basis to expand the physical correlator and they are related in a straightforward way to analytic functionals. This construction also generalizes to any 1D CFT with global symmetry and also multiple correlator cases. Let us discuss the construction of these analytic functional in the context of correlators of four scalar operators transforming under cubic global symmetry. In this case, there are three sectors- singlet(S), symmetric (T), traceless diagonal (V)  where parity even operators are exchanged and in the antisymmetric (A) sector, the parity odd operator gets exchanged. Also the crossing equation leads to four independent sum rules. The analytic functional can be constructed by integrating these crossing equations with appropriate kernel as we discussed above. But here we focus on the construction of these functionals from exchange and contact Witten diagram. The parity even sector can be expanded in terms of scalar exchange Witten diagrams with scalar external legs and the parity odd sector can be expanded in terms of spin-1 exchange Witten diagram with scalar external operators. Then we should add all the regge bounded contact Witten diagrams. In this case, there are three contact Witten diagrams. This construction will result into the following sum rules in cubic anisotropic case,
\begin{equation} \label{cubicfunc}
   \begin{split}
       & \sum_{\Delta} a^S_{\Delta} [\mathcal{N}(\Delta,\Delta_{\phi})(b^s_n(\Delta,\Delta_{\phi})+\frac{2}{N}b^t_{2n}(\Delta,\Delta_{\phi}))]+\frac{2(N-1)}{N}\sum_{\Delta}a^T_{\Delta} \mathcal{N}(\Delta,\Delta_{\phi})b^t_{2n}(\Delta,\Delta_{\phi})\\
       &+\frac{2}{N^2}\sum_{\Delta}a^V_{\Delta}\mathcal{N}(\Delta,\Delta_{\phi})b^t_{2n}(\Delta,\Delta_{\phi}) +2(1-\frac{1}{N})\sum_{\Delta}a^A_{\Delta}b^t_{2n,1}(\Delta,\Delta_{\phi})+\lambda_1 b_{2n,c}^{1,S}+\lambda_2 b_{2n,c}^{2,S}+\lambda_3 b_{2n,c}^{3,S}=0,\\
       & \sum_{\Delta} a^T_{\Delta} [\mathcal{N}(\Delta,\Delta_{\phi})(b^s_n(\Delta,\Delta_{\phi})+b^t_{2n}(\Delta,\Delta_{\phi}))]+\sum_{\Delta}a^S_{\Delta} \mathcal{N}(\Delta,\Delta_{\phi})b^t_{2n}(\Delta,\Delta_{\phi})\\
       &-\frac{1}{N}\sum_{\Delta}a^V_{\Delta}\mathcal{N}(\Delta,\Delta_{\phi})b^t_{2n}(\Delta,\Delta_{\phi}) -\sum_{\Delta}a^A_{\Delta}b^t_{2n,1}(\Delta,\Delta_{\phi})+\lambda_1 b_{2n,c}^{1,T}+\lambda_2 b_{2n,c}^{2,T}+\lambda_3 b_{2n,c}^{3,T}=0
       \end{split}
       \end{equation}
       \begin{equation}
           \begin{split}
                & \sum_{\Delta} a^V_{\Delta} [\mathcal{N}(\Delta,\Delta_{\phi})(b^s_n(\Delta,\Delta_{\phi})+2(1-\frac{1}{N})b^t_{2n}(\Delta,\Delta_{\phi}))]+2\sum_{\Delta}a^S_{\Delta} \mathcal{N}(\Delta,\Delta_{\phi})b^t_{2n}(\Delta,\Delta_{\phi})\\
    &-2\sum_{\Delta}a^T_{\Delta}\mathcal{N}(\Delta,\Delta_{\phi})b^t_{2n}(\Delta,\Delta_{\phi}) -2\sum_{\Delta}a^A_{\Delta}b^t_{2n,1}(\Delta,\Delta_{\phi})+\lambda_1 b_{2n,c}^{1,V}+\lambda_2 b_{2n,c}^{2,V}+\lambda_3 b_{2n,c}^{3,V}=0,\\
    & \sum_{\Delta} a^A_{\Delta} (b^s_{n,1}(\Delta,\Delta_{\phi})+b^t_{2n+1,1}(\Delta,\Delta_{\phi}))+\sum_{\Delta}a^S_{\Delta} \mathcal{N}(\Delta,\Delta_{\phi})b^t_{2n+1}(\Delta,\Delta_{\phi})\\
       &-\frac{1}{N}\sum_{\Delta}a^V_{\Delta}\mathcal{N}(\Delta,\Delta_{\phi})b^t_{2n+1}(\Delta,\Delta_{\phi}) -\sum_{\Delta}a^T_{\Delta}b^t_{2n+1}(\Delta,\Delta_{\phi})+\lambda_1 b_{2n+1,c}^{1,A}+\lambda_2 b_{2n+1,c}^{2,A}+\lambda_3 b_{2n+1,c}^{3,A}=0,
   \end{split}
\end{equation}

We get another set of four equations by replacing $b$ with $a$. Explicit expressions of these quantities are given in appendix \ref{app:expression}. The contact terms are provided in appendix \ref{app:contact}.Note we get sum rules which are labelled by discrete non-negative integer $n$. As we discussed in \eqref{fixinglambda} here also the unknown quantities $\lambda_1$, $\lambda_2$, and $\lambda_3$ are determined by using three  equations and therefore we will miss three functionals here. In our construction, we select the first equation of \eqref{cubicfunc} by setting $n$ equal to $0$ and $1$, as well as setting $n$ equal to $0$ in the second equation of \eqref{cubicfunc}. This essentially results in a finite number of shifts in the orthogonality relations satisfied by these analytic functionals.
\section{Expressions of analytic functionals}\label{app:expression}
Here we provide explicit expressions of the quantities that appeared in our functional sum rules. The purpose of this appendix is to provide all the ingredients that is required to compute CFT data reported in this paper and also to perform  numerical exploration with the analytic functionals. We will upload all the relevant files in online platform in near future. These quantities can be derived following the prescription of \cite{Zhou:2018sfz}. In the case of identical scalars without any global symmetry only the parity even exchanges are allowed and therefore the relevant quantities are related to SL(2,R) block decomposition coefficient of exchange scalar Witten diagram and these expressions appeared in \cite{Zhou:2018sfz} and we repeat below for completeness.
The $s$-channel exchange scalar Witten diagram takes the following form,
    \begin{equation}
    W^s_{\Delta,0}(z)=G_{\Delta}(z)+\sum_n a^s_{n}(\Delta,\Delta_{\phi}) G_{\Delta_{n}}+\sum_n b^s_{n}(\Delta,\Delta_{\phi})  \partial G_{\Delta_{n}},
\end{equation}
where $\Delta_n=2\Delta_{\phi}+2n$ and the decomposition coefficients take the following form,
\begin{equation}
    \begin{split}
     &   b^s_n(\D,\Df)=\frac{\sqrt{\pi } (-1)^{2 n} \Gamma (n+\Delta \phi )^4 \Gamma \left(n+2 \Delta \phi -\frac{1}{2}\right)^2}{(n!)^2 \Gamma (\Delta \phi )^4 (-\Delta +2 \Delta \phi +2 n) (\Delta +2 \Delta \phi +2 n-1) \Gamma (2 (n+\Delta \phi )) \Gamma \left(2 n+2 \Delta \phi -\frac{1}{2}\right)},\\
        & a^s_n(\D,\Df)=\sqrt{\pi } (-1)^{2 n+1} \Gamma (n+\Delta \phi )^4 \Gamma(n+2 \Delta \phi -\frac{1}{2})^2 \bigg(-1+4 n +4 \Delta \phi +(-\Delta +2 \Delta \phi +2 n) \\
        & (\Delta +2 \Delta \phi +2 n-1) (-2 H_{n+\Delta \phi -1}-H_{n+2 \Delta \phi -\frac{3}{2}}+2 H_{4 n+4 \Delta \phi -2}+H_n-\log (4))\bigg).
    \end{split}
\end{equation}
The $t$-channel exchange scalar Witten diagram can be decomposed as,
\begin{equation}
    W^t_{\Delta,0}(z)=\sum_n a^t_{n}(\Delta,\Delta_{\phi}) G_{2\Delta_{\phi}+n}(z)+\sum_n b^t_{n}(\Delta,\Delta_{\phi})\partial G_{2\Delta_{\phi}+n}(z),
\end{equation}
and they can be found by solving the following recursion relations,
for even $n$,
\begin{equation}
    \rho_{n-1}b^t_{n-1}(\Delta,\Delta_{\phi})+\nu_n b^t_n(\Delta,\Delta_{\phi})+\mu_{n+1}b^t_{n+1}(\Delta,\Delta_{\phi})=R_{\frac{n}{2}},
\end{equation}

and for odd $n$,
\begin{equation}
    \rho_{n-1}b^t_{n-1}(\Delta,\Delta_{\phi})+\nu_n b^t_n(\Delta,\Delta_{\phi})+\mu_{n+1}b^t_{n+1}(\Delta,\Delta_{\phi})=0.
\end{equation}
Similarly for $a^t_n(\Delta,\Delta_{\phi})$ even and odd $n$  coefficients follow the following recursions respectively,
\begin{equation}
    \rho_{n-1}a^t_{n-1}(\Delta,\Delta_{\phi})+\nu_n a^t_n(\Delta,\Delta_{\phi})+\mu_{n+1}a^t_{n+1}(\Delta,\Delta_{\phi})+\rho'_{n-1} b^t_{n-1}(\Delta,\Delta_{\phi})+\nu'_n b^t_n+\mu'_{n+1} b^t_{n+1}(\Delta,\Delta_{\phi}) =S_{\frac{n}{2}},
\end{equation}

\begin{equation}
    \rho_{n-1}a^t_{n-1}(\Delta,\Delta_{\phi})+\nu_n a^t_n(\Delta,\Delta_{\phi})+\mu_{n+1}a^t_{n+1}(\Delta,\Delta_{\phi})+\rho'_{n-1} b^t_{n-1}(\Delta,\Delta_{\phi})+\nu'_n b^t_n(\Delta,\Delta_{\phi})+\mu'_{n+1} b^t_{n+1} (\Delta,\Delta_{\phi})=0.
\end{equation}
The $\mu_n$,$\nu_n$, $\rho_n$ has the following form,
\begin{equation}
\begin{split}
   & \mu_n=-n^2\\
    & \nu_n=(\Delta -1) \Delta +\Delta \phi +\frac{1}{2} n (4 \Delta_\phi +n-1)\\
    & \rho_n=-\frac{(2 \Delta_\phi +n)^2 (4 \Delta_\phi +n-1)^2}{4 (4 \Delta_\phi +2 n-1) (4 \Delta_\phi +2 n+1)}
    \end{split}
\end{equation}
and $\mu'_n$,$\nu'_n$, $\rho'_n$ are derivatives of  corresponding expressions with respect to $n$. 
$R_n$ and $S_n$ takes the following form,
\begin{equation}
\begin{split}
    & R_n=\frac{\sqrt{\pi } (-1)^{1-2 n} \Gamma (n+\Delta \phi )^4 \Gamma \left(n+2 \Delta \phi -\frac{1}{2}\right)^2}{(n!)^2 \Gamma (\Delta \phi )^4 \Gamma (2 (n+\Delta \phi )) \Gamma \left(2 (n+\Delta \phi )-\frac{1}{2}\right)},\\
    & S_n=\frac{\sqrt{\pi } (-1)^{-2 n} \Gamma (n+\Delta \phi )^4 \Gamma \left(n+2 \Delta \phi -\frac{1}{2}\right)^2}{(n!)^2 \Gamma (\Delta \phi )^4 \Gamma (2 (n+\Delta \phi )) \Gamma \left(2 (n+\Delta \phi )-\frac{1}{2}\right)}\\
    & \left(2 H_{4 n+4 \Delta \phi -2}-2 H_{n+\Delta \phi -1}-H_{n+2 \Delta \phi -\frac{3}{2}}+H_n-\log (4)\right).
    \end{split}
\end{equation}
In presence of global symmetry the parity odd exchanges appear as well and therefore the relevant quantities are related to SL(2,R) block decomposition coefficient of exchange spin-1 Witten diagram in $AdS_2$. This was constructed following \cite{Zhou:2018sfz} and used before in \cite{Ghosh:2021ruh}. We give the explicit expressions below. The exchange $s$- channel Witten diagram can be decomposed as follows,
\begin{equation}
    W^s_{\Delta,1}(z)=G_{\Delta}(z)+\sum_n a^s_{n,1} G_{\Delta_{n,1}}(z)+\sum_n b^s_{n,1} \partial G_{\Delta_{n,1}}(z),
\end{equation}
where $\Delta_{n,1}=2\Delta_{\phi}+2n+1$ and 

    \begin{equation}
\begin{split}
  &  b^s_{n,1}=\frac{\Gamma \left(\frac{\Delta +1}{2}\right)^2 \Gamma \left(-\frac{\Delta }{2}+\Delta_\phi +\frac{1}{2}\right)^2 \Gamma \left(\frac{\Delta }{2}+\Delta_\phi \right)^2}{\Gamma \left(\frac{\Delta -1}{2}\right)}\\
    & \frac{\Delta  \Gamma \left(\frac{\Delta }{2}\right) \Gamma \left(\Delta +\frac{1}{2}\right) (\Delta \phi +n) 2^{\Delta -2 \Delta_\phi -2 n+1} \Gamma (n+\Delta \phi )^2 \Gamma (n+\Delta_\phi +1) \Gamma \left(n+2 \Delta_\phi +\frac{1}{2}\right)^2}{\Gamma (n+1)^2 (-\Delta +2 \Delta_\phi +2 n+1) (\Delta +2 \Delta \phi +2 n) \Gamma \left(n+\Delta_\phi +\frac{3}{2}\right) \Gamma \left(2 n+2 \Delta_\phi +\frac{1}{2}\right)}
    \end{split}
\end{equation}
{\small
\begin{equation}
\begin{split}
  &  a^s_{n,1}=\frac{\Gamma \left(\frac{\Delta +1}{2}\right)^2 \Gamma \left(-\frac{\Delta }{2}+\Delta_\phi +\frac{1}{2}\right)^2 \Gamma \left(\frac{\Delta }{2}+\Delta_\phi \right)^2}{\Gamma \left(\frac{\Delta -1}{2}\right)}\\
  &\left(\Gamma \left(\frac{\Delta }{2}+1\right) \Gamma \left(\Delta +\frac{1}{2}\right) 2^{\Delta +4 (\Delta_\phi +n)+2} \Gamma (n+\Delta_\phi )^2 \Gamma (n+\Delta_\phi +1)^2 \Gamma \left(n+2 \Delta_\phi +\frac{1}{2}\right)^2\right.\\
  & \left. ((1-\Delta ) \Delta -8 (\Delta_\phi +n)^2 (2 \Delta_\phi +2 n+1)+2 (\Delta_\phi +n) (2 \Delta_\phi +2 n+1) (\Delta -2 \Delta_\phi -2 n-1) (\Delta +2 \Delta_\phi +2 n) \right.\\
  & \left. \frac{-2 H_{n+\Delta_\phi -1}+2 H_{4 (n+\Delta_\phi )}-H_{n+2 \Delta_\phi -\frac{1}{2}}+H_n-\frac{1}{\Delta_\phi +n}-\log (4)}{\pi  (2 \Delta_\phi +2 n+1)^2 \Gamma (n+1)^2 (-\Delta +2 \Delta_\phi +2 n+1)^2 (\Delta +2 \Delta_\phi +2 n)^2 \Gamma (4 n+4 \Delta_\phi +1)})\right)
    \end{split}
\end{equation}}
The $t$-channel exchange spin-1 Witten diagram can be decomposed as follows,
\begin{equation}
    W^t_{\Delta,1}(z)=\sum_n a^t_{n} G_{2\Delta_{\phi}+n}+\sum_n b^t_{n} \partial G_{2\Delta_{\phi}+n},
\end{equation}
and the decomposition coefficients can be found using the following recursion relations,
For even $n$,
\begin{equation}
    \rho_{n-1}b^t_{n-1,1}(\Delta,\Delta_{\phi})+\nu_n b^t_{n,1}(\Delta,\Delta_{\phi})+\mu_{n+1}b^t_{n+1,1}(\Delta,\Delta_{\phi})=R^e_{\frac{n}{2}},
\end{equation}
and for odd $n$,
\begin{equation}
    \rho_{n-1}b^t_{n-1,1}(\Delta,\Delta_{\phi})+\nu_n b^t_{n,1}(\Delta,\Delta_{\phi})+\mu_{n+1}b^t_{{n+1},1}(\Delta,\Delta_{\phi})=R^o_{\frac{n-1}{2}}.
\end{equation}
Similarly for $a^t_n(\Delta,\Delta_{\phi})$, for even n,
\begin{equation}
\begin{split}
    &\rho_{n-1}a^t_{n-1,1}(\Delta,\Delta_{\phi})+\nu_n a^t_{n,1}(\Delta,\Delta_{\phi})+\mu_{n+1}a^t_{n+1,1}(\Delta,\Delta_{\phi})+\rho'_{n-1} b^t_{n-1,1}(\Delta,\Delta_{\phi})\\
    &+\nu'_n b^t_{n,1}(\Delta,\Delta_{\phi})+\mu'_{n+1} b^t_{n+1,1}(\Delta,\Delta_{\phi}) =S^e_{\frac{n}{2}},
    \end{split}
\end{equation}
and for odd n,
\begin{equation}
\begin{split}
    &\rho_{n-1}a^t_{n-1,1}(\Delta,\Delta_{\phi})+\nu_n a^t_{n,1}(\Delta,\Delta_{\phi})+\mu_{n+1}a^t_{n+1,1}(\Delta,\Delta_{\phi})+\rho'_{n-1} b^t_{n-1,1}(\Delta,\Delta_{\phi})\\
    &+\nu'_n b^t_{n,1}(\Delta,\Delta_{\phi})+\mu'_{n+1} b^t_{n+1,1}(\Delta,\Delta_{\phi})
    =S^o_{\frac{n-1}{2}}.
    \end{split}
\end{equation}
$R_n^e$ and $R_n^o$ stands for even and odd contribution respectively. The source term appearing in the right hand side of the above equations take the following explicit form,
\begin{equation}
    \begin{split}
        & R^e_n=\frac{(\Delta -1) (-1)^{2 n} \Gamma (2 \Delta +1) 2^{-2 (\Delta +\Delta_\phi +n+1)} \Gamma (n+\Delta_\phi )^3 \Gamma \left(n+2 \Delta_\phi -\frac{1}{2}\right)^2}{\Gamma \left(\frac{\Delta +1}{2}\right)^4 \Gamma (n+1)^2 \Gamma \left(-\frac{\Delta }{2}+\Delta_\phi +\frac{1}{2}\right)^2 \Gamma \left(\frac{\Delta }{2}+\Delta_\phi \right)^2}\\
        & \bigg(\frac{2 \pi  \left(-(1-4 \Delta_\phi )^2 \Delta_\phi +4 n^3+4 (5 \Delta_\phi -1) n^2-4 \Delta_\phi  n+n\right)}{\Gamma \left(n+\Delta_\phi +\frac{1}{2}\right) \Gamma \left(2 n+2 \Delta_\phi +\frac{1}{2}\right)}\\
        &-\frac{n 64^{\Delta_\phi +n} \left(2 (1-4 \Delta_\phi )^2+14 n^2+(46 \Delta_\phi -11) n\right) \Gamma (n+\Delta_\phi )}{\Gamma (4 (n+\Delta_\phi ))}\bigg),\\
        & R^o_n=\frac{(-1)^{2 n+1} \Gamma \left(\frac{\Delta }{2}+1\right) \Gamma \left(\Delta +\frac{1}{2}\right) 2^{\Delta +4 \Delta_\phi +4 n} \Gamma (n+\Delta_\phi )^2 \Gamma (n+\Delta_\phi +1)^2 \Gamma \left(n+2 \Delta_\phi +\frac{1}{2}\right)^2}{\pi  \Gamma \left(\frac{\Delta -1}{2}\right) \Gamma \left(\frac{\Delta +1}{2}\right)^2 (2 \Delta_\phi +2 n+1) \Gamma (n+1)^2 \Gamma \left(-\frac{\Delta }{2}+\Delta_\phi +\frac{1}{2}\right)^2 \Gamma \left(\frac{\Delta }{2}+\Delta_\phi \right)^2 \Gamma (4 (n+\Delta_\phi ))},
        \end{split}
        \end{equation}
\begin{equation}
\begin{split}
    & S^e_n=\frac{\Gamma \left(\Delta +\frac{1}{2}\right) 2^{\Delta -2 \Delta_\phi -2 n-6} \Gamma (n+\Delta_\phi )^3 \Gamma \left(n+2 \Delta_\phi -\frac{1}{2}\right)^2}{\pi  (n!)^2 \Gamma \left(\frac{\Delta -1}{2}\right) \Gamma \left(\frac{\Delta +1}{2}\right)^2 \Gamma \left(-\frac{\Delta }{2}+\Delta_\phi +\frac{1}{2}\right)^2 \Gamma \left(\frac{\Delta }{2}+\Delta_\phi \right)^2}\\
        & \bigg[ \frac{64 \pi  \Delta  \Gamma \left(\frac{\Delta }{2}\right) \left(\Delta_\phi  (4 \Delta_\phi -1)+6 n^2+3 (4 \Delta \phi -1) n\right) (\psi ^{(0)}(2 (n+\Delta \phi ))-2 \psi ^{(0)}(n+\Delta \phi ))}{\Gamma \left(n+\Delta_\phi +\frac{1}{2}\right) \Gamma \left(2 n+2 \Delta_\phi -\frac{1}{2}\right)}\\
       &+\frac{\Gamma \left(\frac{\Delta }{2}+1\right) 64^{\Delta \phi +n} \Gamma (n+\Delta_\phi -1)}{n (2 \Delta \phi +2 n-1) (4 \Delta_\phi +2 n-1) (4 \Delta_\phi +4 n-3) (4 \Delta_\phi +4 n-1) \Gamma (4 (n+\Delta \phi -1))}\\
       &\bigg(48 n^4+48 (4 \Delta_\phi -1) n^3+6 (4 \Delta_\phi  (16 \Delta_\phi -7)+3) n^2+3 (8 \Delta_\phi -1) (1-4 \Delta_\phi )^2 n\\
        & +2 n (4 \Delta_\phi +2 n-1) (4 \Delta \phi +4 n-1) \left(\Delta_\phi  (4 \Delta_\phi -1)+6 n^2+3 (4 \Delta_\phi -1) n\right) \left\{-\psi ^{(0)}\left(n+2 \Delta_\phi +\frac{1}{2}\right) \right.\\
        & \left. +\psi ^{(0)}\left(2 n+2 \Delta \phi +\frac{1}{2}\right)+\psi ^{(0)}(n)\right\}\bigg)\bigg],\\
        & S^o_n=\frac{\Delta  \Gamma \left(\frac{\Delta }{2}\right) \Gamma \left(\Delta +\frac{1}{2}\right) 2^{\Delta +4 \Delta_\phi +4 n} \Gamma (n+\Delta_\phi )^2 \Gamma (n+\Delta_\phi +1)^2 \Gamma \left(n+2 \Delta_\phi +\frac{1}{2}\right)^2}{\pi  \Gamma \left(\frac{\Delta -1}{2}\right) \Gamma \left(\frac{\Delta +1}{2}\right)^2 (2 \Delta_\phi +2 n+1)^2 \Gamma (n+1)^2 \Gamma \left(-\frac{\Delta }{2}+\Delta_\phi +\frac{1}{2}\right)^2 \Gamma \left(\frac{\Delta }{2}+\Delta_\phi \right)^2 \Gamma (4 n+4 \Delta_\phi +1)}\\
        & [-4 \Delta_\phi +2 (\Delta_\phi +n) (2 \Delta_\phi +2 n+1) \left(-2 H_{n+\Delta_\phi -1}+2 H_{4 (n+\Delta_\phi )}-H_{n+2 \Delta_\phi -\frac{1}{2}}+H_n-\log (4)\right)-4 n-3].
    \end{split}
\end{equation}

The above recursion relations can be solved once the $b^t_0(\Delta,\Delta_{\phi})$,$a^t_0(\Delta,\Delta_{\phi})$,$b^t_{0,1}(\Delta,\Delta_{\phi})$,$a^t_{0,1}(\Delta,\Delta_{\phi})$ are given and we also use the fact that $b^t_{-1}(\Delta,\Delta_{\phi})$=$a^t_{-1}(\Delta,\Delta_{\phi})$=$b^t_{-1,1}(\Delta,\Delta_{\phi})$=$a^t_{-1,1}(\Delta,\Delta_{\phi})$=0. The explicit expressions of these quantities are given in the attached notebook. 

\section{Contact terms} \label{app:contact}
Here we write down three explicit forms of contact terms that appear in the construction of Polyakov blocks relevant for cubic anisotropic group. It is easier to write them in terms of Mellin variables. There are two contact terms that we can write down that involve no derivatives and essentially correspond to two different contractions that are possible in this theory. This gives rise to the following two contact terms:
\begin{equation}
   C^1=\{D(z),0,N  D(z),0\},\,\,\,\,\,  C^2=\{0, D(z),-N D(z),0\},
\end{equation}
where 
\begin{equation}
   D(z)=\int [ds][dt] z^s(1-z)^{2t} \Gamma(\Delta_{\phi}-s)^2\Gamma(-t)^2\Gamma(s+t)^2.
\end{equation}
The other contact term involves four fields with two derivatives acting on them. This takes the following form, 
\begin{equation}
   C^{3,j}(z)=\int [ds][dt] z^s(1-z)^{2t} M^i(s,t)\Gamma(\Delta_{\phi}-s)^2\Gamma(-t)^2\Gamma(s+t)^2,
\end{equation}
where j denotes the different irreps of cubic symmetry group and the Mellin amplitudes are given by,
\begin{equation}
\begin{split}
    M(s)=&\bigg\{ \left(1-\frac{2}{N}\right) \left(\Delta \phi -\frac{s}{2}\right)_1-\left(\left(\Delta \phi -\frac{t}{2}\right)_1+\left(\Delta \phi -\frac{u}{2}\right)_1\right)\\
    & -\left(\Delta \phi -\frac{s}{2}\right)_1, -2 \left(\Delta \phi -\frac{s}{2}\right)_1,\left(\Delta \phi -\frac{t}{2}\right)_1-\left(\Delta \phi -\frac{u}{2}\right)_1 \bigg\}.
   \end{split}
\end{equation}
From the Mellin transformation it is a straightforward task to write down the following decomposition of the first two contact diagrams
\begin{equation}
  C^{i,j}(z)=\sum_n b^{i,j}_{2n,c} G_{\Delta_n}(z)  +\sum_n a^{i,j}_{2n,c} G_{\Delta_n}(z)  
\end{equation}
and the third diagram can be decomposed as,
\begin{equation}
    C^{3,j}(z)=\sum_n b^{3,j}_{n,c} \partial G_{2\Delta_{\phi}+n}(z)  +\sum_n a^{3,j}_{n,c} G_{2\Delta_{\phi}+n}(z).
\end{equation}
Here $i$ can be either 1 or 2 denoting the first two contact diagrams and again $j$ index runs over all four irreducible sectors of the cubic anisotropic group.

\bibliography{defect}
\bibliographystyle{jhep}

\end{document}